\documentclass[%
reprint,
superscriptaddress,
showpacs,preprintnumbers,
amsmath,amssymb,
prb,
]{revtex4-2}

\usepackage{graphicx}
\usepackage{dcolumn}
\usepackage{xfrac}
\usepackage{bm}
\usepackage{color}
\usepackage[export]{adjustbox}
\usepackage{xr-hyper}
\usepackage{hyperref}
\usepackage{tabularx}

\usepackage{amsmath}
\usepackage{amsfonts}
\usepackage{amssymb}
\usepackage{tabularx}
\usepackage{chemformula}
\usepackage{soul,xcolor}
\setstcolor{red}

\newcommand{\ub}{$\mu_{\text B}$}

\begin{document}

\title{Two types of colossal magnetoresistance with distinct mechanisms in \ch{Eu5In2As6}}  

\author{Sudhaman~Balguri$^\dagger$}
\affiliation{Department of Physics, Boston College, Chestnut Hill, MA 02467, USA}
\thanks{These authors contributed equally to this work.}

\author{Mira~B.~Mahendru$^\dagger$}
\affiliation{Department of Physics, Boston College, Chestnut Hill, MA 02467, USA}

\author{Enrique~O.~González~Delgado}
\affiliation{Department of Physics, Boston College, Chestnut Hill, MA 02467, USA}

\author{Kyle~Fruhling}
\affiliation{Department of Physics, Boston College, Chestnut Hill, MA 02467, USA}

\author{Xiaohan~Yao}
\affiliation{Department of Physics, Boston College, Chestnut Hill, MA 02467, USA}

\author{David~E.~Graf}
\affiliation{National High Magnetic Field Laboratory, Tallahassee, Florida 32310, USA}

\author{Jose~A.~Rodriguez-Rivera}
\affiliation{NIST Center for Neutron Research, National Institute of Standards and Technology, Gaithersburg, Maryland 20899, USA}
\affiliation{Department of Materials Science and Eng., University of Maryland, College Park, MD 20742-2115}

\author{Adam~A.~Aczel}
\affiliation{Neutron Scattering Division, Oak Ridge National Laboratory, Oak Ridge, Tennessee 37831, USA}

\author{Andreas~Rydh}
\affiliation{Department of Physics, Stockholm University, 106 91 Stockholm, Sweden}

\author{Jonathan~Gaudet}
\affiliation{NIST Center for Neutron Research, National Institute of Standards and Technology, Gaithersburg, Maryland 20899, USA}
\affiliation{Department of Materials Science and Eng., University of Maryland, College Park, MD 20742-2115}

\author{Fazel~Tafti}
\email{fazel.tafti@bc.edu}
\affiliation{Department of Physics, Boston College, Chestnut Hill, MA 02467, USA}


\begin{abstract}
Recent reports of colossal negative magnetoresistance (CMR) in a few magnetic semimetals and semiconductors have attracted attention, because these materials are devoid of the conventional mechanisms of CMR such as mixed valence, double exchange interaction, and Jahn-Teller distortion.
New mechanisms have thus been proposed, including topological band structure, ferromagnetic clusters, orbital currents, and charge ordering.
The CMR in these compounds has been reported in two forms: either a resistivity peak or a resistivity upturn suppressed by a magnetic field.
Here we reveal both types of CMR in a single antiferromagnetic semiconductor \ch{Eu5In2As6}.
Using the transport and thermodynamic measurements, we demonstrate that the peak-type CMR is likely due to the percolation of magnetic polarons with increasing magnetic field, while the upturn-type CMR is proposed to result from the melting of a charge order under the magnetic field.
We argue that similar mechanisms operate in other compounds, offering a unifying framework to understand CMR in seemingly different materials. 
\end{abstract}

\maketitle


\section{\label{sec:introduction}Introduction}
Antiferromagnetic (AFM) materials with coupled charge, spin, and lattice degrees of freedom are coming to the forefront of electronic, spintronic, and caloritronic applications~\cite{baltz_antiferromagnetic_2018,han_coherent_2023,xu_observation_2022}.
The Eu-based Zintl compounds are emerging as a fascinating family of functional AFM semiconductors with intertwined degrees of freedom.
They exhibit exotic properties such as fluctuating Weyl nodes in \ch{EuCd2As2}~\cite{ma_spin_2019,jo_manipulating_2020}, axion insulator phase in \ch{EuIn2As2}~\cite{xu_higher-order_2019,riberolles_magnetic_2021}, and colossal magnetoresistance (CMR) in \ch{EuCd2P2}~\cite{wang_colossal_2021,sunko_spin-carrier_2023,homes_optical_2023} and \ch{Eu5In2Sb6}~\cite{rosa_colossal_2020,ghosh_colossal_2022,morano_noncollinear_2024,souza_microscopic_2022,rahn_unusual_2023}.
Since CMR is particularly applicable in sensing and logic devices, it has been sought in a variety of materials including the chalcogenide Zintl phase \ch{Mn3Si2Te6}~\cite{zhang_control_2022} and Dirac semimetals such as CeSbTe and EuAuSb~\cite{singha_colossal_2023,ram_magnetotransport_2024}.  
The general features of CMR in these materials are similar; they all exhibit a peak in their zero-field resistivity curves $\rho(T)$ which is suppressed by an external magnetic field.
In some cases, such as \ch{Eu5In2Sb6} and \ch{Mn3Si2Te6}, a sharp upturn is observed in $\rho(T)$ at low temperatures which is also suppressed by the magnetic field.
Different mechanisms have been proposed for either type of CMR (peak or upturn), including ferromagnetic (FM) clusters~\cite{sunko_spin-carrier_2023}, band structure reconstruction~\cite{zhang_electronic_2023}, chiral orbital currents~\cite{zhang_control_2022}, and interplay between Dirac fermions and charge ordering~\cite{singha_colossal_2023}. 

Here, we reveal both types of CMR in a single AFM semiconductor \ch{Eu5In2As6} and discuss the underlying mechanism for each type.
A few properties of the title compound make it ideal for understanding CMR in the context of other related materials mentioned above.
First, \ch{Eu5In2As6} is topologically trivial as shown by recent first-principles calculations~\cite{varnava_engineering_2022}, so Dirac fermions are irrelevant to its CMR, unlike the proposals for CeSbTe and EuAuSb~\cite{singha_colossal_2023,ram_magnetotransport_2024}.
Second, \ch{Eu5In2As6} is a semiconductor with a band gap of 35~meV corresponding to temperature and field scales of about 400~K and 300~T.
Thus, its sharp resistivity upturn at 15.5~K which is completely suppressed by 9~T, as shown here, cannot be due to a field- or temperature-induced transition from an insulator to a metal.
Third, Eu$^{2+}$ ions in \ch{Eu5In2As6} are not at the center of high-symmetry anionic polyhedra, so they are not subjected to orbital currents as proposed for \ch{Mn3Si2Te6}~\cite{zhang_control_2022}. 

In this work, we argue that the peak-like CMR in \ch{Eu5In2As6} at high temperatures is related to polaron physics while the upturn-like CMR at low temperatures is potentially related to charge segregation or charge ordering.
The extremely large magnitudes of both effects indicate a strong coupling among charge, spin, and lattice degrees of freedom, a property that could be common among the Zintl compounds mentioned above.

\section{\label{sec:methods}Methods}
\begin{figure*}
  \includegraphics[width=\textwidth]{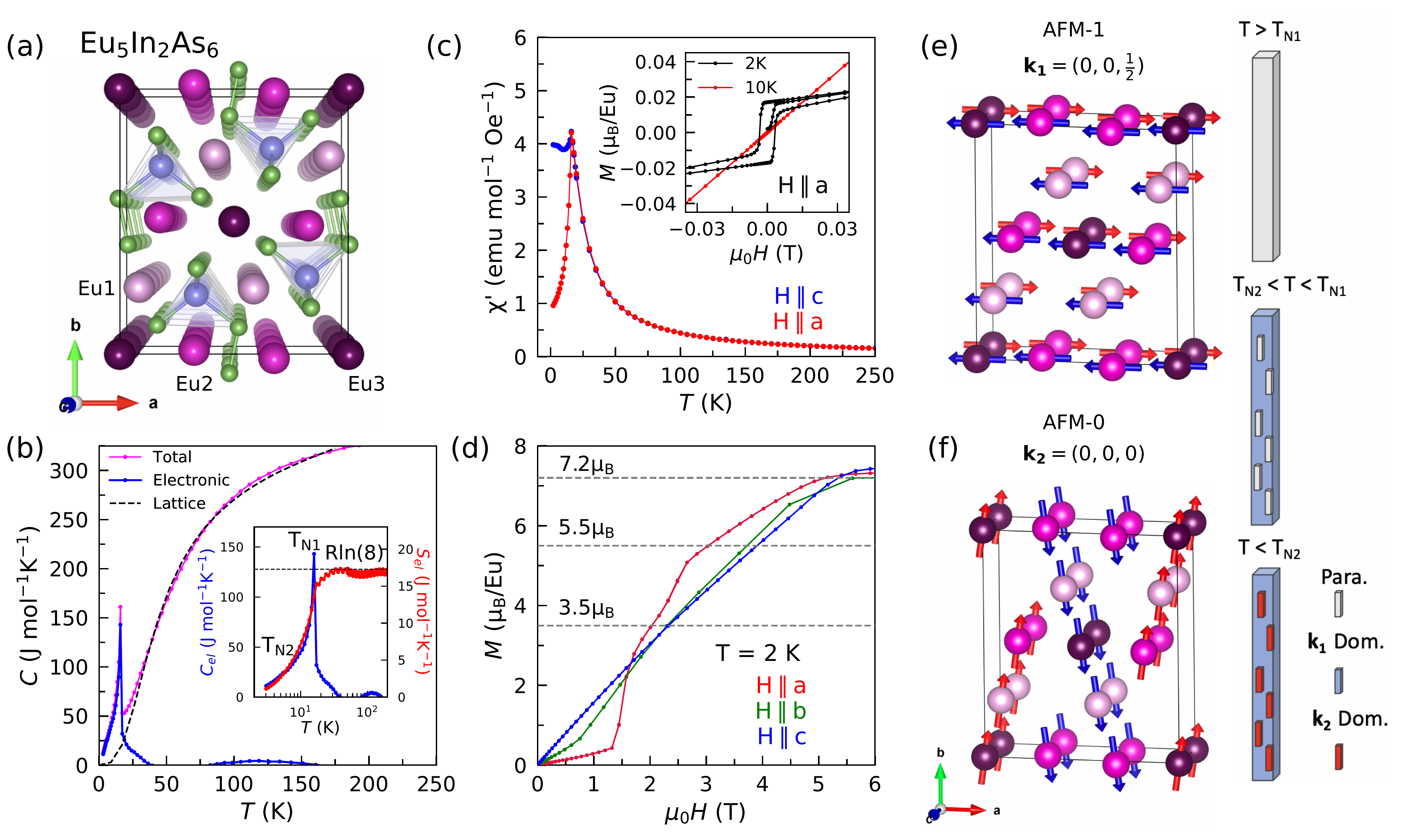}
  \caption{\label{fig:STRUCTURE}
 (a) Orthorhombic unit cell of \ch{Eu5In2As6} in the space group $Pbam$.
 (b) Total, electronic, and lattice heat capacity shown in magenta, blue, and black colors.
 The lattice background is modeled assuming anharmonic Debye model~\cite{stern_theory_1958,suppmatt} with $\Theta_\text{D}=196$~K.
 Inset shows the electronic heat capacity (blue) and entropy (red).
 (c) AC magnetic susceptibility curves suggest AFM ordering with in-plane moments and alternating direction along the $c$-axis.
 The hysteresis loop at $T<T_\text{N2}$ in the inset suggests FM ordering along the $a$-axis.
 (d) Three steps in the $M(H)$ curve when $H\|a$ suggest spin flops on different Eu sites.
 (e) The AFM spin structure determined by neutron diffraction in subsection~\ref{subsec:Neutron} at $T<T_\text{N1}$.
 The ordered moments are FM-aligned along the $a$-axis and AFM-aligned along the $c$-axis.   The AFM-1 structure persists to zero temperature.
 (f) At $T<T_\text{N2}$, new magnetic domains with $\mathbf{k_2}=(0,0,0)$ emerge in separate regions of the sample. 
 According to the analysis of panel (b), these domains (AFM-0) constitute less than 20\% of the sample volume.
 The diagram on the right shows the phase separation scenario where the sample volume is represented by a rectangular prism.
 The 3 different prisms sketch the spatial variation of magnetic domains assuming the three zero-field magnetic phases that we observed.
 The gray color represents paramagnetic (para.) domains, while the blue and red domains respectively correspond to $\mathbf{k_1}$ (panel (e)) and $\mathbf{k_2}$ domains (panel (f)). 
  }
\end{figure*}
Polycrystalline samples of \ch{Eu5In2As6} have been previously synthesized~\cite{tomitaka_bipolar_2021,radzieowski_divalent_2020} but single crystals were not available until now.
We report the first crystal growth of this compound here (Supplemental Material~\cite{suppmatt}).
Powder x-ray diffraction was performed using a Bruker D8 ECO instrument. 
The FullProf suite~\cite{rodriguez-carvajal_recent_1993} and VESTA software~\cite{momma_vesta_2011} were used for the Rietveld refinement and crystal visualization.
Physical properties were measured using Quantum Design Dynacool-PPMS and MPMS-3.

Neutron diffraction data were collected using the thermal neutron triple-axis spectrometer VERITAS at the High Flux Isotope Reactor (HFIR) in Oak Ridge National Laboratory. 
A few crystals were mounted on an Al plate with either the $c$-axis or $b$-axis perpendicular to the scattering plane to probe the (H,K,0) or (H,0,L) Bragg peaks, respectively. 
Representation analysis was performed using SARAh~\cite{wills_new_2000}. 
The Cooper-Nathans formalism was used to calculate the resolution function of VERITAS~\cite{chesser_derivation_1973}, and the Wuesch-Prewitt algorithm~\cite{wuensch_corrections_1965} was used to calculate the angle-dependent neutron transmission of the sample.
The statistical error for each intensity point corresponds to 1 standard deviation.

\section{\label{sec:results}Results and Discussion}

\subsection{\label{subsec:Structure}Structural Analysis}
\begin{figure*}
  \includegraphics[width=\textwidth]{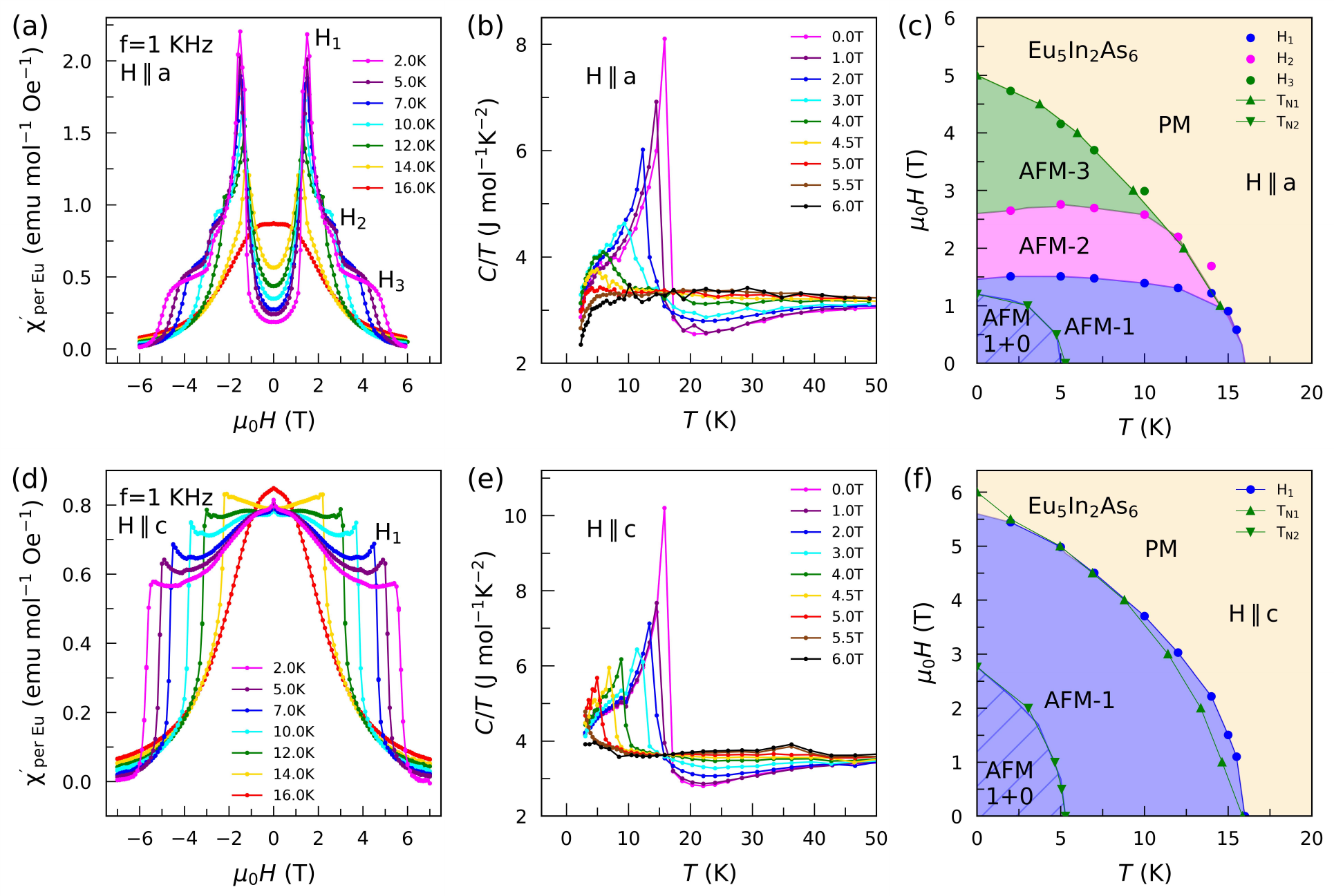}
  \caption{\label{fig:MAG}
 (a) Field dependence of the real part of the AC susceptibility $\chi'(H)$ at several temperatures. 
 The $H_1$, $H_2$, and $H_3$ features imply spin flop transitions. 
 They correspond to the three steps in DC magnetization in the red curve on Fig.~\ref{fig:STRUCTURE}d.
 (b) Temperature dependence of the heat capacity at several magnetic fields.
 The sharp peak marks $T_{\text{N1}}$.
 (c) The magnetic phase diagram when $H\|a$.
 The area marked by striped lines indicates the coexistence of AFM-0 and AFM-1 domains (Figs.~\ref{fig:STRUCTURE}e,f) in separate regions of the sample.
 (d,e,f) Similar data and phase diagram to (a,b,c) when $H\|c$.
  }
\end{figure*}
From a chemical standpoint, \ch{Eu5In2As6} is considered a Zintl compound, i.e. it is an intermetallic material that exhibits semiconducting behavior due to the charge balance between large cationic and anionic complexes that make up its lattice structure~\cite{kauzlarich_zintl_2023}.  
In this compound, the positive charge of [Eu$_5$]$^{10+}$ is balanced by the negative charges of [In$_2$As$_4$]$^{6-}$ and [As$_2$]$^{4-}$ complexes~\cite{childs_five_2019}. 

\ch{Eu5In2As6} has an orthorhombic unit cell with three inequivalent Eu sites and a characteristic chain stacking of atoms as shown in Fig.~\ref{fig:STRUCTURE}a.
Details of the x-ray refinement are presented in the supplemental Fig.~S1 and Tables~S1 and S2~\cite{suppmatt}.
The crystal structure of \ch{Eu5In2As6} belongs to the same non-symmorphic space group, $Pbam$ (No. 55), as its sister compounds, \ch{Ba5In2Sb6} and \ch{Eu5In2Sb6}, which are predicted to be topological hourglass and axion insulators, respectively~\cite{wieder_wallpaper_2018,rosa_colossal_2020}. 

A recent density functional theory (DFT) calculation shows that while \ch{Eu5In2Sb6} and \ch{Eu5In2Bi6} could be topological, \ch{Eu5In2As6} is a trivial semiconductor~\cite{varnava_engineering_2022}.
We specifically chose this compound to demonstrate that a large CMR does not rely on topological aspects of the band structure.
This is important because the salient features of CMR in the trivial semiconductor \ch{Eu5In2As6} are similar to those in topological systems such as the Dirac semimetals CeSbTe and EuAuSb~\cite{singha_colossal_2023,ram_magnetotransport_2024}, Weyl semimetal \ch{EuCd2As2}~\cite{ma_spin_2019}, and axion insulator \ch{Eu5In2Sb6}~\cite{rahn_unusual_2023}.

Another exotic mechanism proposed for CMR is the modifictaion of chiral orbital currents (COC) in \ch{Mn3Si2Te6} with applied field~\cite{zhang_control_2022}.
Central to this mechanism is the octahedral coordination of six Te atoms around each Mn$^{2+}$ ion, enabling the interaction between the COC and Mn$^{2+}$ spin.
As seen in Fig.~\ref{fig:STRUCTURE}a, the Eu$^{2+}$ ions in \ch{Eu5In2As6} are not directly surrounded by a high-symmetry polyhedron of In and As.
Therefore, COC is not a viable mechanism for the CMR in \ch{Eu5In2As6}.

The above structural analysis dismisses topology and COC as the source of CMR in \ch{Eu5In2As6}.
Next, we will discuss its magnetic structure and phase diagram, before a detailed analysis of CMR.

\subsection{\label{subsec:Magnetism}Magnetic Transitions} 
The heat capacity data in Fig.~\ref{fig:STRUCTURE}b reveal two AFM transitions in zero-field at $T_{\text{N1}}=16.0(2)$~K and $T_{\text{N2}}=5.4(3)$~K.
Magnetism is driven by Eu$^{2+}$ ions in the half-filled $4f^{7}$ configuration ($^{8}S_{7/2}$) with $L=0$ and $S=7/2$, consistent with the saturated entropy $R\ln(8)$~Jmol$^{-1}$K$^{-1}$ in the inset of Fig.~\ref{fig:STRUCTURE}b, saturated magnetization 7.2~\ub\ in Fig.~\ref{fig:STRUCTURE}d, and effective moment $\mu_{\text{eff}}=7.8(6)$~\ub\ from a Curie-Weiss analysis in Fig.~S2.

The inset of Fig.~\ref{fig:STRUCTURE}b shows that approximately 10\% of the full entropy is released by $T_\text{N2}$, 70\% is released between $T_\text{N2}$ and $T_\text{N1}$, and 20\% is released above $T_\text{N1}$.
For comparison, \ch{Eu5In2Sb6} also exhibits two transitions at $T_{\text{N1}}=14$~K and $T_{\text{N2}}=7$~K but it releases nearly equal amounts of entropy at each transition~\cite{morano_noncollinear_2024}.
In both materials, 20\% of total entropy is released above $T_\text{N1}$, well into the paramagnetic (PM) phase, suggesting short-range magnetic correlations (polarons) that survive up to $3\,T_\text{N1}$~\cite{ale_crivillero_magnetic_2023,crivillero_surface_2022}. 

Temperature dependence of the in-plane and out-of-plane magnetic susceptibility in Fig.~\ref{fig:STRUCTURE}c suggests an AFM order with the moments in the $ab$ plane but alterning along the $c$-axis.
The inset of Fig.~\ref{fig:STRUCTURE}c shows the absence of a hysteresis loop at $T_\text{N2}<T<T_\text{N1}$, so the order is purely AFM in this temperature range.
However, a small hysteresis loop opens below $T_{\text N2}$ when $H\|a$, suggesting a finite in-plane FM component when $T<T_\text{N2}$.

The behavior of $M(H)$ curves along three different crystallographic directions is shown in Fig.~\ref{fig:STRUCTURE}d.
When $H\|c$, the $M(H)$ curve is strictly linear, indicating an AFM order along the $c$-axis.
In contrast, the $M(H)$ curve exhibits three steps when $H\|a$, indicating consecutive spin flops on different Eu sites within the $ab$-planes.
The steps are less apparent when $H\|b$, suggesting that the $a$-axis is the easy axis.
A positive Curie-Weiss temperature ($\Theta_\text{CW}=+15$~K) in Fig.~S2 indicates FM correlations despite an AFM order, consistent with FM correlations within the $ab$-planes in spite of AFM ordering along the $c$-axis.

The spin structure at zero-field is discussed in subsection~\ref{subsec:Neutron}. 
While we cannot rule-out the possibility that both magnetic ordering at $T_\text{N1}$ and $T_\text{N2}$ form within the same magnetic domain, we suggest an alternative scenario where spatially separated magnetic domains form independently below each transition temperature.
In this scenario, the majority domains (more than $80\%$ of the sample volume according to the entropy release in Fig.~\ref{fig:STRUCTURE}b) form below $T_\text{N1}$ with a spin structure shown in Fig.~\ref{fig:STRUCTURE}e.
This structure is consistent with the $\chi(T)$ and $M(H)$ curves shown in Figs.~\ref{fig:STRUCTURE}c,d.
Specifically, it shows FM spins within the $ab$-planes with an easy $a$-axis, and with alternating directions along the $c$-axis.
Less than 20$\%$ of the sample volume remains paramagnetic down to $T_\text{N2}$ where it orders into the structure shown in Fig.~\ref{fig:STRUCTURE}f, which has an FM component along the $a$-axis and explains the hysteresis observed in the inset of Fig.~\ref{fig:STRUCTURE}c at 2~K. 
Ferromagnetism along the $a$-axis was also observed in the sister compound \ch{Eu5In2Sb6}, but is only stabilized for $T_\text{N2}<T<T_\text{N1}$~\cite{rosa_colossal_2020,morano_noncollinear_2024,rahn_unusual_2023}. 

\subsection{\label{subsec:Phase}Phase Diagrams}
\begin{figure*}
  \includegraphics[width=\textwidth]{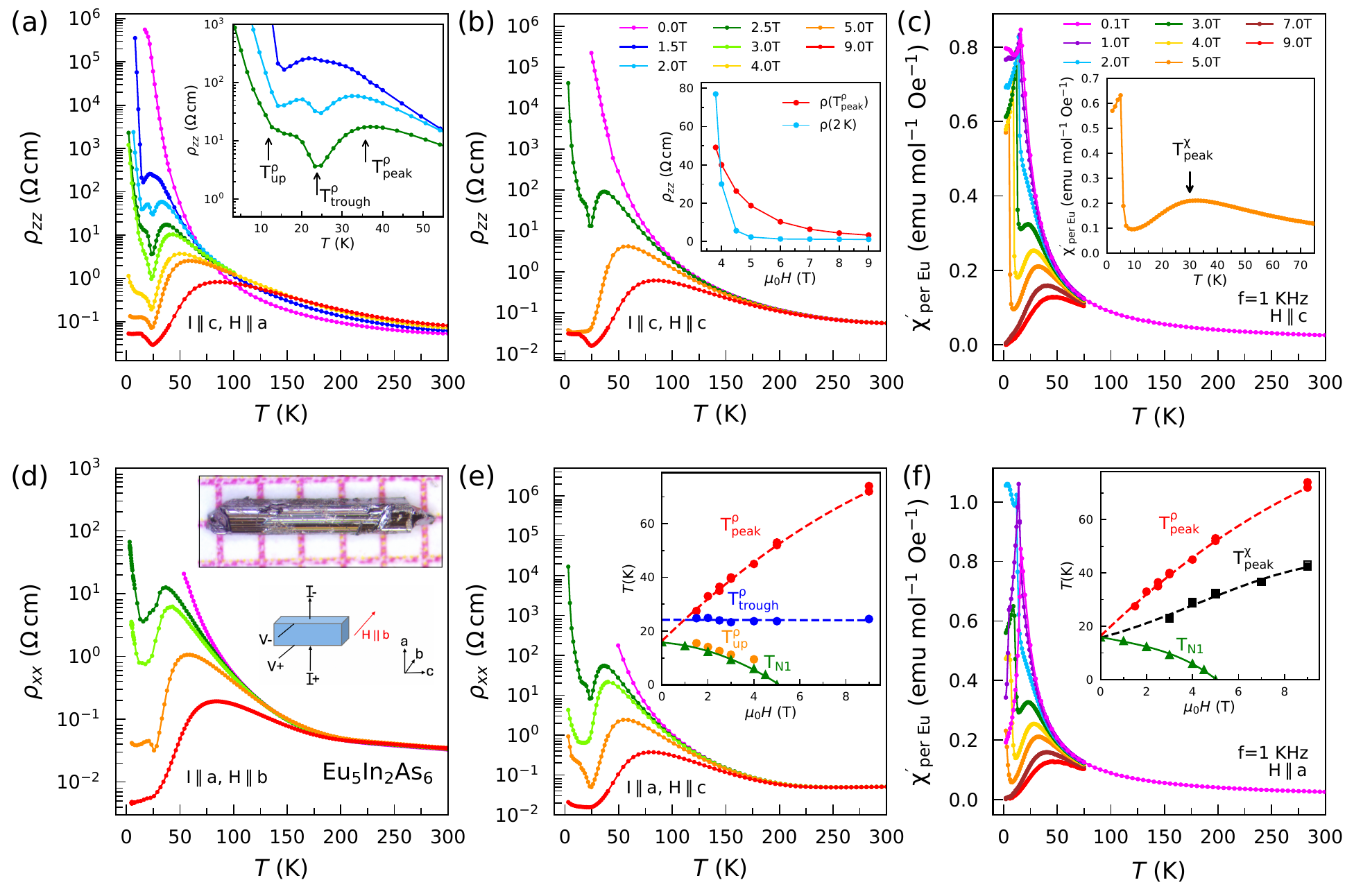}
  \caption{\label{fig:RESIST}
 (a) Resistivity curves measured with current and voltage leads along the $c$-axis ($\rho_{zz}$) at different fields with $H\|a$.
 At zero field, $\rho_{zz}$ exceeds the detection limit of our instrument below 20~K.
 The inset magnifies $\rho_{zz}(T)$ to indicate $T^\rho_\text{peak}$, $T^\rho_\text{trough}$ and $T^\rho_\text{up}$.
 (b) $\rho_{zz}(T)$ measured with $H\|c$ shows similar behavior as with $H\|a$.
 The inset highlights the different field dependence between the two types of CMR, namely the peak-type CMR ($\rho(T^\rho_\text{peak})$) and upturn-type CMR ($\rho(2~\text{K})$).
 The $\rho_{zz}(T)$ data from which this plot is constructed are shown in Fig.~S10.
 (c) The broad peak in the AC susceptibility data at $T>T_\text{N1}$ is evidence of polarons. The sharp peak is due to magnetic ordering.
 (d) Resistivity curves measured with current and field along the $a$-axis ($\rho_{xx}$) with $H\|b$.
 The inset shows a crystal of \ch{Eu5In2As6} with the longest dimension being along the $c$-axis.
 (e) $\rho_{xx}(T)$ curves with $H\|c$.
 The inset compares the field dependence of $T^\rho_\text{peak}$, $T^\rho_\text{up}$, and $T_\text{N1}$.
 (f) AC susceptibility data with $H\|a$ shows similar behavior as with $H\|c$.
 The inset compares the field dependence of $T^\rho_\text{peak}$ (circles), $T^\chi_\text{peak}$ (squares), and $T_\text{N1}$ (triangles).
  }
\end{figure*}
We used detailed measurements of the AC susceptibility and nanocalorimetry to map the temperature-field phase diagram of \ch{Eu5In2As6}.
Figure~\ref{fig:MAG}a shows three characteristic fields ($H_1$, $H_2$, and $H_3$) in the AC susceptibility data at $T=2$~K.
They coincide with the three steps in the $M(H\|a)$ curve in Fig.~\ref{fig:STRUCTURE}d.
With increasing temperature, these characteristic fields become smaller and merge as shown in the phase diagram of Fig.~\ref{fig:MAG}c.
The triangular data points that define the outer boundary of the $H\|a$ phase diagram indicate the field-dependence of $T_\text{N1}$ in the heat capacity data (Fig.~\ref{fig:MAG}b).
The $T_\text{N2}$ phase boundary is determined via a low-temperature nanocalorimetry technique~\cite{tagliati_differential_2012} shown in Fig.~S3 (see also the supplemental tunnel diode oscillation data in Figs.~S4 and S5).
The striped region on the phase diagrams of Fig.~\ref{fig:MAG}c indicates the phase separation between AFM-0 and AFM-1 spin structures in different regions of the sample with AFM-1 being the majority domain.

The AC susceptibility and heat capacity data for $H\|c$ in Figs.~\ref{fig:MAG}d,e are used to map the phase diagram with out-of-plane field in Fig.~\ref{fig:MAG}f.
Supplemental nanocalorimetry data in Fig.~S3 are used to define the $T_\text{N2}$ boundary when $H\|c$.
The rich phase diagrams of Figs.~\ref{fig:MAG}c,f suggest that \ch{Eu5In2As6} is a multi-critical system~\cite{wang_topological_2021} with several magnetic states accessible by modest magnetic fields. 
This is the result of having three inequivalent Eu sites in a non-symmorphic crystal structure (Figs.~\ref{fig:STRUCTURE}a,e,f) that enables asymmetric exchange interactions~\cite{wang_topological_2021}.

\subsection{\label{subsec:CMR}Colossal Magnetoresistance (CMR)} 
The resistivity data reveal two types of CMR in \ch{Eu5In2As6}, one at low temperatures near $T_\text{N1}$ associated with a sharp upturn in the resistivity, and another at high temperatures well above $T_\text{N1}$ associated with a broad peak in the resistivity.
The resistivity upturn and resistivity peak are affected differently by the magnetic field, leading to two distinct types of CMR.
Figure~\ref{fig:RESIST}a shows the $\rho_{zz}(T)$ curves obtained with both current and voltage leads along the crystallographic $c$-axis.
The pink curve is measured at zero field, and the rest of the curves are obtained under different magnetic fields labeled by different colors in the legend of Fig.~\ref{fig:RESIST}b.
For consistency, we follow the same color code to specify the magnetic field in Figs.~\ref{fig:RESIST}a,b,d,e.

The blue curve in Fig.~\ref{fig:RESIST}a shows that at 1.5~T, $\rho_{zz}(T)$ increases as the temperature is decreased from 300~K and forms a broad peak at around 27~K, before it exhibits a sharp upturn at 15.5~K. 
Looking closely at the blue curve (inset of Fig.~\ref{fig:RESIST}a), one could identify a shallow trough near 24~K between the peak at 27~K and  the upturn at 15.5~K.
The shallow trough becomes deeper with increasing field but its characteristic temperature does not change.
These three features in the resistivity curves, namely the broad peak at $T^\rho_\text{peak}$, the trough at $T^\rho_\text{trough}$, and the upturn at $T^\rho_\text{up}$, evolve with increasing magnetic field as shown in the inset of Fig.~\ref{fig:RESIST}e.

The suppression of the resistivity peak and resistivity upturn in Fig.~\ref{fig:RESIST}a leads to two distinct types of CMR as the field is increased from zero to 9~T.
The high-temperature peak-type CMR at $T^\rho_\text{peak}$ spans 3 orders of magnitude while its characteristic temperature $T^\rho_\text{peak}$ increases with increasing field (inset of Fig.~\ref{fig:RESIST}e).
The low-temperature upturn-type CMR is even larger, spanning 7 orders of magnitude.
Changing the magnetic field direction from $H\|a$ (Fig.~\ref{fig:RESIST}a) to $H\|c$ (Fig.~\ref{fig:RESIST}b) does not affect either type of CMR. 
Changing the electric current direction also leaves both types of CMR qualitatively unchanged as seen in the $\rho_{xx}(T)$ curves with $H\|b$ (Fig.~\ref{fig:RESIST}d) and $H\|c$ (Fig.~\ref{fig:RESIST}e).
Inset of Fig.~\ref{fig:RESIST}b reveals a gradual suppression of the resistivity peak at $T^\rho_\text{peak}$ in contrast to the rapid drop in the resistivity upturn at 2~K.
Note that the drop in the resistivity upturn is complete by 5.5~T, the field at which the AFM order is fully suppressed (Figs.~\ref{fig:MAG}c,f), whereas the suppression of resistivity peak extends to 9~T. 
Since the peak-type and upturn-type CMRs are marked by different resistivity features, different temperature scales, and different field dependencies, they must have different origins.
Below, we will discuss each type in more detail.

\emph{Peak-type CMR.} The broad peak in the resistivity curves at $T^\rho_\text{peak}$ could be attributed to magnetic polarons, which are nanometer size FM clusters of Eu$^{2+}$ moments coupled via conduction electrons~\cite{pohlit_evidence_2018,das_magnetically_2012,sullow_magnetotransport_2000,zhang_nonlinear_2009,murakawa_giant_2023}. 
When the sample is cooled from room temperature to $T^\rho_\text{peak}$, polarons proliferate and grow in size, so they scatter electrons at a higher rate and increase the resistivity.
With further decreasing temperature below $T^\rho_\text{peak}$, polarons overlap and provide percolation paths for conduction electrons, so the resistivity decreases.
Increasing the magnetic field makes it easier to polarize the FM clusters, so the polaron peak at $T^\rho_\text{peak}$ shifts to higher temperatures with increasing field, as seen in Figs.~\ref{fig:RESIST}a,b,d,e.

The inset of Fig.~\ref{fig:RESIST}e highlights the opposite effect of the magnetic field on magnetic polarons and magnetic order.
The red circles mark $T^\rho_\text{peak}$ in $\rho_{zz}(T)$ curves at different fields, while the green triangles mark $T_\text{N1}$ from heat capacity data in Fig.~\ref{fig:MAG}e.
With increasing field, $T^\rho_\text{peak}$ is shifted to higher temperatures because local moments polarize more easily and fluctuate less under an external field.
In contrast, $T_\text{N1}$ is shifted to lower temperatures since the AFM order is suppressed by the field.

We also find evidence of magnetic polarons in the AC susceptibility data in Figs.~\ref{fig:RESIST}c and \ref{fig:RESIST}f for $H\|c$ and $H\|a$, respectively.
Since Eu $f$-moments are localized, a Curie-Weiss behavior ($\chi'\propto 1/T$) is expected at $T> T_\text{N1}$.
However, Figs.~\ref{fig:RESIST}c and \ref{fig:RESIST}f show a deviation from the $1/T$ behavior with a broad peak at $T^\chi_\text{peak}$ (inset of Fig.~\ref{fig:RESIST}c).
The broad peak at $T^\chi_\text{peak}$ suggests a reduction of the magnetic susceptibility due to short-range magnetic correlations developing within magnetic polarons. 

In both susceptibility and resistivity data, the broad polaronic peaks are shifted to higher temperatures with increasing magnetic field, as shown in the inset of Fig.~\ref{fig:RESIST}f.
However, the susceptibility peaks occur at lower temperatures compared to the resistivity peaks ($T^\chi_\text{peak}<T^\rho_\text{peak}$ in Fig.~\ref{fig:RESIST}f).
This is because the susceptibility and resistivity reveal two different aspects of polarons.
Susceptibility measures the spin correlations within polarons that develop at lower temperatures, while resistivity measures the scattering of conduction electrons by polarons that starts from higher temperatures.
Such a difference between transport and thermodynamic quantities has also been reported in \ch{EuB6}, the archetypal polaronic material~\cite{manna_lattice_2014,chatterjee_spin-polaron_2004}.

\emph{Upturn-type CMR.} The second type of CMR is marked by the 7 orders-of-magnitude suppression of the resistivity upturn at the lowest measured temperature (2~K) as the field increases from zero to 9~T.
The inset of Fig.~\ref{fig:RESIST}a shows that the resistivity upturn initiates at $T^\rho_\text{trough}$ but then it accelerates by the onset of magnetic ordering at $T^\rho_\text{up}$ which coincides with $T_\text{N1}$ (inset of Fig.~\ref{fig:RESIST}e).
At $T<T^\rho_\text{up}$, resistivity continues to increase down to the lowest measured temperatures.
As the magnetic order becomes weaker with increasing field, the resistivity upturn also attenuates (Figs.~\ref{fig:RESIST}a,b,d,e).
Specifically, the inset of Fig.~\ref{fig:RESIST}b shows a near complete suppression of the resistivity upturn by 5.5~T, where the AFM order is fully suppressed.
This is consistent with the parallel suppression of $T^\rho_\text{up}$ and $T_\text{N1}$ with increasing field from zero to 5.5~T in the inset of Fig.~\ref{fig:RESIST}e.

The parallel suppression of the upturn-type CMR and magnetic ordering with increasing field may suggest that the upturn-type CMR is caused by magnetic ordering.
However, magnetic ordering typically reduces electrical resistance by removing spin fluctuations as a source of scattering.
Thus, we can propose two potential mechanisms for the observed upturn-type CMR.
The first possibility is a reconstruction of the electronic structure by the AFM order, i.e. a metal-insulator transition induced by the Slater mechanism~\cite{slater_magnetic_1951}.
The second possibility is a charge ordering concomitant with AFM ordering, a scenario which is well established in manganite perovoskites with CMR~\cite{tomioka_magnetic-field-induced_1997,tomioka_collapse_1995,rao_charge_2000}.
The first scenario is challenged by DFT calculations that suggest \ch{Eu5In2As6} is a narrow gap semiconductor, so it already has a small gap before entering the AFM phase~\cite{varnava_engineering_2022}. 
The second scenario is challenged by the fact that unlike manganites, \ch{Eu5In2As6} is not a Mott insulator, so its physics could be different from those materials.
Future ARPES experiments could reveal the band reconstruction~\cite{li_colossal,zhang_electronic_2023} while inelastic x-ray scattering~\cite{tomioka_magnetic-field-induced_1997} and microwave impedance microscopy~\cite{lai_mesoscopic_2010} could reveal charge ordering.

It is worth noting that the temperature scale $T^\rho_\text{trough}$, where the resistivity upturn initiates, is field independent as shown in the inset of Fig.~\ref{fig:RESIST}e.
Notably, this feature in resistivity shows no corresponding thermodynamic signature either in magnetization or heat capacity measurements. 
Thus, the origin of this feature remains unclear and requires further investigation.

\subsection{\label{subsec:Neutron}Neutron Diffraction}
\begin{figure}
  \includegraphics[width=0.5\textwidth]{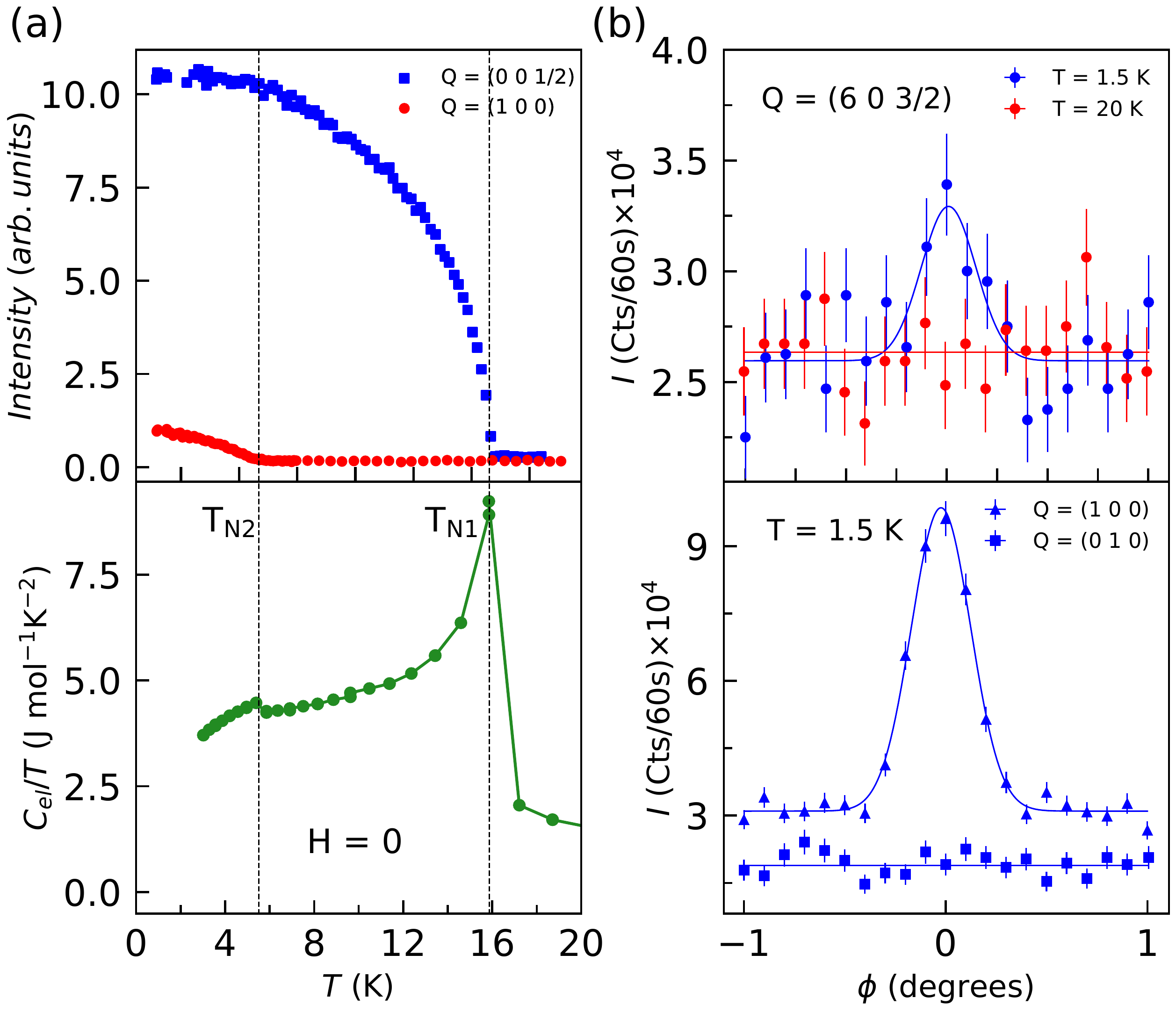}
  \caption{\label{fig:NEUTRON}
 (a) Temperature dependence of the neutron magnetic Bragg peak intensities at $\mathbf{Q}=(0,0,\frac{1}{2})$ and $\mathbf{Q}=(1,0,0)$ in the top panel is compared to the temperature dependence of heat capacity in the bottom panel. 
 (b) The top panel shows the intensity of $\mathbf{Q}=(6,0,\frac{3}{2})$ peaks at both $T=1.5$~K and $T=20$~K.
 The bottom panel shows that magnetic Bragg scattering is present for $\mathbf{Q}=(1,0,0)$ (triangles) but absent for $\mathbf{Q}=(0,1,0)$ (squares) at $T=1.5$~K. 
 }
\end{figure}
We used neutron diffraction data to resolve the spin structure of the ordered state below $T_\text{N1}$ and $T_\text{N2}$ in \ch{Eu5In2As6} at zero field.
Figure~\ref{fig:NEUTRON}a shows the onset of magnetic diffraction peaks indexed by a propagation vector $\mathbf{k_1}=(0,0,\frac{1}{2})$ below $T_\text{N1}$ and another set of peaks indexed by $\mathbf{k_2}=(0,0,0)$ below $T_\text{N2}$.
Given the likelihood of a charge ordering in \ch{Eu5In2As6}, we associated this phenomenology to a magnetic phase segregation scenario where $\mathbf{k_1}$ and $\mathbf{k_2}$ magnetic domains form separately within the sample. 
This is consistent with the lack of intensity changes observed at $\mathbf{Q}=(0,0,\frac{1}{2})$ when the $k_2$ peaks onsets below $T_\text{N2}$.
The $\mathbf{k_1}$ wave vector corresponds to an inter-layer AFM order along the $c$-axis, which persists to zero temperature, while $\mathbf{k_2}$ corresponds to the development of a spin structure with an FM inter-layer coupling arising below $T_\text{N2}$.
Note that the peak intensity of $\mathbf{k_2}$ is about 10 times smaller than $\mathbf{k_1}$, indicating that the $\mathbf{k_2}$ domains form the majority of the sample volume.
For comparison, \ch{Eu5In2Sb6} shows a reverse order of wave vectors, such that $\mathbf{k_2}$ and $\mathbf{k_1}$ appear with comparable intensities below $T_\text{N1}$ and $T_\text{N2}$, respectively~\cite{morano_noncollinear_2024}.

Following the symmetry analysis of Refs.~\cite{morano_noncollinear_2024,rahn_unusual_2023} for \ch{Eu5In2Sb6}, we find 8 irreducible representations (irreps) for the \ch{Eu5In2As6} space group $Pbam$ compatible with the propagation vectors $\mathbf{k_1}=(0,0,\frac{1}{2})$ and $\mathbf{k_2}=(0,0,0)$. 
However, only 4 irreps ($\Gamma_1,\Gamma_3,\Gamma_5,\Gamma_7$) can produce magnetization on all three Eu Wyckoff sites -- a condition imposed by the full entropy of $R\ln8$ in Fig.~\ref{fig:STRUCTURE}b.
These four irreps correspond to an AFM or FM arrangement of Eu$^{2+}$ spins within each $ab$-plane oriented along either the $a$ ($\Gamma_3$, $\Gamma_5$), $b$ ($\Gamma_3$, $\Gamma_5$), or $c$ ($\Gamma_1$, $\Gamma_7$) directions. 
The second-order phase transitions revealed by heat capacity (Fig.~\ref{fig:STRUCTURE}b) and order parameter measurements (Fig.~\ref{fig:NEUTRON}a) require a single irrep to explain the magnetic structure.

We first analyze the $\mathbf{k_1}$ magnetic domains that onset below $T_\text{N1}$ and persist to zero temperature with at least 70\% volume fraction according to the entropy analysis in Fig.~\ref{fig:STRUCTURE}b.
As determined by the order parameter curve collected at $\mathbf{Q}~=~(0,0,\frac{1}{2})$ (Fig.\ref{fig:NEUTRON}a), magnetic Bragg peaks for these domains are observed at $\mathbf{Q}~=~(0,0,L)$.
Since neutron scattering only probes magnetization perpendicular to the momentum transfer $\mathbf{Q}$, any irrep with spins strictly along the $c$-direction ($\Gamma_1$ and $\Gamma_7$) are excluded.
This leaves $\Gamma_3$ and $\Gamma_5$ as the only viable choices.

$\Gamma_3$ has an FM basis vector with the spins pointing along the $a$-axis ($\Gamma_3(\psi_1)$), and an AFM basis vector with the spins pointing along the $b$-axis ($\Gamma_3(\psi_2)$). 
$\Gamma_5$ has the opposite spin arrangement, i.e. it has an FM basis vector with spins along the $b$-axis ($\Gamma_5(\psi_2)$), and an AFM basis vector with spins along the $a$-axis ($\Gamma_5(\psi_1)$).
An FM order within the $ab$ plane produces magnetic Bragg peaks at $\mathbf{Q}~=~(\text{even},0,\frac{L}{2})$, while an AFM order produces them at $\mathbf{Q}~=~(\text{odd},0,\frac{L}{2})$. 
The top panel of Fig.~\ref{fig:NEUTRON}b shows magnetic Bragg scattering at even positions, indicating an FM arrangement within the $ab$-plane. 
Therefore, the spin structure can be represented by either $\Gamma_3(\psi_1)$ or $\Gamma_5(\psi_2)$.  
A quantitative refinement of the Bragg peak intensities could distinguish between $\Gamma_3(\psi_1)$ and $\Gamma_5(\psi_2)$, but this contrast disappears due to the large neutron absorption cross-sections of Eu. 
However, the $\Gamma_3(\psi_1)$ basis vector is consistent with the low-temperature magnetization of \ch{Eu5In2As6} (Fig.\ref{fig:STRUCTURE}d), which shows spin-flop transitions for field applied along the $a$-axis. 
This contrasts with the magnetization for a field applied along the $b$-axis where such spin-flop transitions are not observed and would be expected to occur for the $\Gamma_5(\psi_2)$ structure.
Therefore, we conclude that $\Gamma_3(\psi_1)$ is the appropriate irrep to describe the magnetic domains forming below $T_\text{N1}$ in \ch{Eu5In2As6}.
This irrep also describes the low-temperature magnetic phase of \ch{Eu5In2Sb6}~\cite{morano_noncollinear_2024,rahn_unusual_2023}.
We depict the resulting $\mathbf{k_1}$ magnetic domains in Fig.~\ref{fig:STRUCTURE}e.

We next turn our attention to the details of the $\mathbf{k_2}$ magnetic domains.
Our bulk magnetization data in Fig.~\ref{fig:STRUCTURE}c shows a hysteresis loop opening below $T_\text{N2}$, indicating a net FM component along the $a$-axis.
We also detected magnetic Bragg peaks at structurally non-allowed Bragg positions, such as $\mathbf{Q}~=~(1,0,0)$ plotted in the bottom panel of Fig.~\ref{fig:NEUTRON}b. 
This indicates that the $\mathbf{k_2}=(0,0,0)$ domains also have an AFM component within the $ab$ plane. 
Since no magnetic Bragg peaks were found at $\mathbf{Q}~=~(0,H,0)$ positions (Fig.~\ref{fig:NEUTRON}b, bottom panel), the spin polarization of this AFM component is along the $b$-axis.
Thus, the magnetic structure below $T_\text{N2}$ must have a net FM component along the $a$-axis and an AFM component along the $b$-axis.
This means the irrep for the $\mathbf{k_2}=(0,0,0)$ domains are identical to the ones for $\mathbf{k_1}=(0,0,\frac{1}{2})$, but the spins do not reverse going along the $c$-direction~\cite{morano_noncollinear_2024,rahn_unusual_2023}. 
The spin structure for the $\mathbf{k_2}$ domains is represented in Fig.~\ref{fig:STRUCTURE}f.

Finally, we note that our neutron diffraction data cannot distinguish a multi-$\mathbf{k}$ structure from our proposed magnetic phase segregation scenario as they both lead to identical zero-field neutron diffraction patterns. 
Magnetic phase segragation between the same AFM-1 and AFM-0 domains was also suggested for \ch{Eu5In2Sb6}~\cite{rahn_unusual_2023}, but in this latter compound the AFM-0 state forms the majority domains.
Furthermore, we propose that a phase separation is quite likely if charge ordering/separation is responsible for the upturn-type CMR.
The competition between charge order and AFM order over the same carriers leads to enhanced scattering of conduction electrons, giving rise to the resistivity upturn.
With increasing field, both charge and AFM order are suppressed, giving rise to the upturn-type CMR.
Future studies using magnetic imaging techniques may reveal magnetic and charge phase separation in \ch{Eu5In2As6}.

For completeness, we also discuss the multi-$\mathbf{k}$ scenario in the supplemental information~\cite{suppmatt}.

\section{\label{sec:summary}Summary and Outlook}
The two types of CMR discussed here have been observed in different Zintl compounds but they were not explained in a unifying manner.
For example, CeSb$_{0.1}$Te$_{1.9}$, \ch{EuAuSb}, and \ch{EuCd2As2} show the peak-type CMR~\cite{singha_colossal_2023,ram_magnetotransport_2024,ma_spin_2019} whereas \ch{EuCd2P2}, \ch{Mn3Si2Te6} and \ch{Eu5In2Sb6} show both the peak-type CMR and the upturn-type CMR~\cite{wang_colossal_2021,zhang_control_2022,rosa_colossal_2020}.
Based on the similarity of transport data across different Zintl compounds, we argue that similar mechanisms are at work to produce the two types of CMR in these materials.
Specifically, we attribute the peak-type CMR to percolation of magnetic polarons and the upturn-type CMR to either melting of a charge ordered state or band structure reconstruction by the AFM order (Slater mechanism).

The CMR induced by magnetic polarons implies a coupling between short-range FM ordering and charge transport, while the CMR associated with charge ordering reflects a coupling between long-range AFM ordering and charge transport. 
Such an intricate interplay between magnetism (both long-range and short-range) and the charge transport is a key feature of many Eu-based Zintl phases~\cite{rahn_coupling_2018,rahn_unusual_2023,sunko_spin-carrier_2023}, potentially driven by the large spin of Eu$^{2+}$ and considerable magnetoelastic coupling. It is noteworthy that a strong influence of the $4f$ states of Europium, through their coupling to valence and conduction states, on the band structure and charge transport of \ch{EuCd2\textit{X}2\ (\textit{X}=P,As,Sb)} systems has already been demonstrated~\cite{li_colossal,nasrallah_magneto-optical_2024}.

An indirect manifestation of magnetoelastic coupling is the dependence of the magnetic state of Eu-based Zintl compounds on the choice of flux in crystal growth~\cite{jo_manipulating_2020,chen_manipulating_2024}. 
These characteristics position Eu-based CMR materials as promising candidates for magnetic sensing, spin valve, and piezoresistive technologies~\cite{ghosh_colossal_2022}.

In the supplemental Figs.~S6 and S7, we present an Arrhenius analysis on the zero-field resistivity data from two samples of \ch{Eu5In2As6} revealing significantly different activation gaps of 35~meV and 125~meV in samples S1 and S2, respectively.
Hall effect data in Fig.~S8 reveal different n-type carrier concentrations in samples S1 ($10^{18}$~cm$^{-3}$) and S2 ($10^{16}$~cm$^{-3}$), so
the sample with a larger gap (S1) shows a smaller carrier concentration.
The different gap values and carrier concentrations in different samples suggest that \ch{Eu5In2As6} is a self-doped semiconductor.
Remarkably, all the interesting transport phenomena in this material, and possibly other Eu-based CMR materials, seem to originate from a small density of extrinsic carriers.

\section*{ACKNOWLEDGMENTS}
F.T. and X.Y. acknowledge funding from the U.S. Department of Energy under Award No. SC0023124.
This material is based upon work supported by the Air Force Office of Scientific Research under award numbers FA2386-21-1-4059 and FA9550-23-1-0124.
A. R. acknowledges support from the Swedish Research Council, D. Nr. 2021-04360.
A portion of this work was performed at the National High Magnetic Field Laboratory, which is supported by the National Science Foundation Cooperative Agreement No. DMR-1644779 and the state of Florida. 
The support for neutron scattering was provided by the Center for High-Resolution Neutron Scattering, a partnership between the National Institute of Standards and Technology and the National Science Foundation under Agreement No. DMR-2010792. 
The identification of any commercial product or trade name does not imply endorsement or recommendation by the National Institute of Standards and Technology. 
A portion of this research used resources at the High Flux Isotope Reactor, a DOE Office of Science User Facility operated by the Oak Ridge National Laboratory. 
The beam time was allocated to VERITAS on proposal number IPTS-31509.




\bibliography{Balguri_Mahendru_12dec2024}

\begin{thebibliography}{52}%
\makeatletter
\providecommand \@ifxundefined [1]{%
 \@ifx{#1\undefined}
}%
\providecommand \@ifnum [1]{%
 \ifnum #1\expandafter \@firstoftwo
 \else \expandafter \@secondoftwo
 \fi
}%
\providecommand \@ifx [1]{%
 \ifx #1\expandafter \@firstoftwo
 \else \expandafter \@secondoftwo
 \fi
}%
\providecommand \natexlab [1]{#1}%
\providecommand \enquote  [1]{``#1''}%
\providecommand \bibnamefont  [1]{#1}%
\providecommand \bibfnamefont [1]{#1}%
\providecommand \citenamefont [1]{#1}%
\providecommand \href@noop [0]{\@secondoftwo}%
\providecommand \href [0]{\begingroup \@sanitize@url \@href}%
\providecommand \@href[1]{\@@startlink{#1}\@@href}%
\providecommand \@@href[1]{\endgroup#1\@@endlink}%
\providecommand \@sanitize@url [0]{\catcode `\\12\catcode `\$12\catcode `\&12\catcode `\#12\catcode `\^12\catcode `\_12\catcode `\%12\relax}%
\providecommand \@@startlink[1]{}%
\providecommand \@@endlink[0]{}%
\providecommand \url  [0]{\begingroup\@sanitize@url \@url }%
\providecommand \@url [1]{\endgroup\@href {#1}{\urlprefix }}%
\providecommand \urlprefix  [0]{URL }%
\providecommand \Eprint [0]{\href }%
\providecommand \doibase [0]{https://doi.org/}%
\providecommand \selectlanguage [0]{\@gobble}%
\providecommand \bibinfo  [0]{\@secondoftwo}%
\providecommand \bibfield  [0]{\@secondoftwo}%
\providecommand \translation [1]{[#1]}%
\providecommand \BibitemOpen [0]{}%
\providecommand \bibitemStop [0]{}%
\providecommand \bibitemNoStop [0]{.\EOS\space}%
\providecommand \EOS [0]{\spacefactor3000\relax}%
\providecommand \BibitemShut  [1]{\csname bibitem#1\endcsname}%
\let\auto@bib@innerbib\@empty
\bibitem [{\citenamefont {Baltz}\ \emph {et~al.}(2018)\citenamefont {Baltz}, \citenamefont {Manchon}, \citenamefont {Tsoi}, \citenamefont {Moriyama}, \citenamefont {Ono},\ and\ \citenamefont {Tserkovnyak}}]{baltz_antiferromagnetic_2018}%
  \BibitemOpen
  \bibfield  {author} {\bibinfo {author} {\bibfnamefont {V.}~\bibnamefont {Baltz}}, \bibinfo {author} {\bibfnamefont {A.}~\bibnamefont {Manchon}}, \bibinfo {author} {\bibfnamefont {M.}~\bibnamefont {Tsoi}}, \bibinfo {author} {\bibfnamefont {T.}~\bibnamefont {Moriyama}}, \bibinfo {author} {\bibfnamefont {T.}~\bibnamefont {Ono}},\ and\ \bibinfo {author} {\bibfnamefont {Y.}~\bibnamefont {Tserkovnyak}},\ }\bibfield  {title} {\bibinfo {title} {Antiferromagnetic spintronics},\ }\href {https://doi.org/10.1103/RevModPhys.90.015005} {\bibfield  {journal} {\bibinfo  {journal} {Reviews of Modern Physics}\ }\textbf {\bibinfo {volume} {90}},\ \bibinfo {pages} {015005} (\bibinfo {year} {2018})}\BibitemShut {NoStop}%
\bibitem [{\citenamefont {Han}\ \emph {et~al.}(2023)\citenamefont {Han}, \citenamefont {Cheng}, \citenamefont {Liu}, \citenamefont {Ohno},\ and\ \citenamefont {Fukami}}]{han_coherent_2023}%
  \BibitemOpen
  \bibfield  {author} {\bibinfo {author} {\bibfnamefont {J.}~\bibnamefont {Han}}, \bibinfo {author} {\bibfnamefont {R.}~\bibnamefont {Cheng}}, \bibinfo {author} {\bibfnamefont {L.}~\bibnamefont {Liu}}, \bibinfo {author} {\bibfnamefont {H.}~\bibnamefont {Ohno}},\ and\ \bibinfo {author} {\bibfnamefont {S.}~\bibnamefont {Fukami}},\ }\bibfield  {title} {\bibinfo {title} {Coherent antiferromagnetic spintronics},\ }\href {https://doi.org/10.1038/s41563-023-01492-6} {\bibfield  {journal} {\bibinfo  {journal} {Nature Materials}\ }\textbf {\bibinfo {volume} {22}},\ \bibinfo {pages} {684} (\bibinfo {year} {2023})}\BibitemShut {NoStop}%
\bibitem [{\citenamefont {Xu}\ \emph {et~al.}(2022)\citenamefont {Xu}, \citenamefont {He}, \citenamefont {Zhou}, \citenamefont {Qu}, \citenamefont {Huang},\ and\ \citenamefont {Chien}}]{xu_observation_2022}%
  \BibitemOpen
  \bibfield  {author} {\bibinfo {author} {\bibfnamefont {J.}~\bibnamefont {Xu}}, \bibinfo {author} {\bibfnamefont {J.}~\bibnamefont {He}}, \bibinfo {author} {\bibfnamefont {J.-S.}\ \bibnamefont {Zhou}}, \bibinfo {author} {\bibfnamefont {D.}~\bibnamefont {Qu}}, \bibinfo {author} {\bibfnamefont {S.-Y.}\ \bibnamefont {Huang}},\ and\ \bibinfo {author} {\bibfnamefont {C.}~\bibnamefont {Chien}},\ }\bibfield  {title} {\bibinfo {title} {Observation of {Vector} {Spin} {Seebeck} {Effect} in a {Noncollinear} {Antiferromagnet}},\ }\href {https://doi.org/10.1103/PhysRevLett.129.117202} {\bibfield  {journal} {\bibinfo  {journal} {Physical Review Letters}\ }\textbf {\bibinfo {volume} {129}},\ \bibinfo {pages} {117202} (\bibinfo {year} {2022})}\BibitemShut {NoStop}%
\bibitem [{\citenamefont {Ma}\ \emph {et~al.}(2019)\citenamefont {Ma}, \citenamefont {Nie}, \citenamefont {Yi}, \citenamefont {Jandke}, \citenamefont {Shang}, \citenamefont {Yao}, \citenamefont {Naamneh}, \citenamefont {Yan}, \citenamefont {Sun}, \citenamefont {Chikina}, \citenamefont {Strocov}, \citenamefont {Medarde}, \citenamefont {Song}, \citenamefont {Xiong}, \citenamefont {Xu}, \citenamefont {Wulfhekel}, \citenamefont {Mesot}, \citenamefont {Reticcioli}, \citenamefont {Franchini}, \citenamefont {Mudry}, \citenamefont {Müller}, \citenamefont {Shi}, \citenamefont {Qian}, \citenamefont {Ding},\ and\ \citenamefont {Shi}}]{ma_spin_2019}%
  \BibitemOpen
  \bibfield  {author} {\bibinfo {author} {\bibfnamefont {J.-Z.}\ \bibnamefont {Ma}}, \bibinfo {author} {\bibfnamefont {S.~M.}\ \bibnamefont {Nie}}, \bibinfo {author} {\bibfnamefont {C.~J.}\ \bibnamefont {Yi}}, \bibinfo {author} {\bibfnamefont {J.}~\bibnamefont {Jandke}}, \bibinfo {author} {\bibfnamefont {T.}~\bibnamefont {Shang}}, \bibinfo {author} {\bibfnamefont {M.~Y.}\ \bibnamefont {Yao}}, \bibinfo {author} {\bibfnamefont {M.}~\bibnamefont {Naamneh}}, \bibinfo {author} {\bibfnamefont {L.~Q.}\ \bibnamefont {Yan}}, \bibinfo {author} {\bibfnamefont {Y.}~\bibnamefont {Sun}}, \bibinfo {author} {\bibfnamefont {A.}~\bibnamefont {Chikina}}, \bibinfo {author} {\bibfnamefont {V.~N.}\ \bibnamefont {Strocov}}, \bibinfo {author} {\bibfnamefont {M.}~\bibnamefont {Medarde}}, \bibinfo {author} {\bibfnamefont {M.}~\bibnamefont {Song}}, \bibinfo {author} {\bibfnamefont {Y.-M.}\ \bibnamefont {Xiong}}, \bibinfo {author} {\bibfnamefont {G.}~\bibnamefont {Xu}}, \bibinfo {author} {\bibfnamefont {W.}~\bibnamefont {Wulfhekel}},
  \bibinfo {author} {\bibfnamefont {J.}~\bibnamefont {Mesot}}, \bibinfo {author} {\bibfnamefont {M.}~\bibnamefont {Reticcioli}}, \bibinfo {author} {\bibfnamefont {C.}~\bibnamefont {Franchini}}, \bibinfo {author} {\bibfnamefont {C.}~\bibnamefont {Mudry}}, \bibinfo {author} {\bibfnamefont {M.}~\bibnamefont {Müller}}, \bibinfo {author} {\bibfnamefont {Y.~G.}\ \bibnamefont {Shi}}, \bibinfo {author} {\bibfnamefont {T.}~\bibnamefont {Qian}}, \bibinfo {author} {\bibfnamefont {H.}~\bibnamefont {Ding}},\ and\ \bibinfo {author} {\bibfnamefont {M.}~\bibnamefont {Shi}},\ }\bibfield  {title} {\bibinfo {title} {Spin fluctuation induced {Weyl} semimetal state in the paramagnetic phase of {EuCd2As2}},\ }\href {https://doi.org/10.1126/sciadv.aaw4718} {\bibfield  {journal} {\bibinfo  {journal} {Science Advances}\ }\textbf {\bibinfo {volume} {5}},\ \bibinfo {pages} {eaaw4718} (\bibinfo {year} {2019})}\BibitemShut {NoStop}%
\bibitem [{\citenamefont {Jo}\ \emph {et~al.}(2020)\citenamefont {Jo}, \citenamefont {Kuthanazhi}, \citenamefont {Wu}, \citenamefont {Timmons}, \citenamefont {Kim}, \citenamefont {Zhou}, \citenamefont {Wang}, \citenamefont {Ueland}, \citenamefont {Palasyuk}, \citenamefont {Ryan}, \citenamefont {McQueeney}, \citenamefont {Lee}, \citenamefont {Schrunk}, \citenamefont {Burkov}, \citenamefont {Prozorov}, \citenamefont {Bud'ko}, \citenamefont {Kaminski},\ and\ \citenamefont {Canfield}}]{jo_manipulating_2020}%
  \BibitemOpen
  \bibfield  {author} {\bibinfo {author} {\bibfnamefont {N.~H.}\ \bibnamefont {Jo}}, \bibinfo {author} {\bibfnamefont {B.}~\bibnamefont {Kuthanazhi}}, \bibinfo {author} {\bibfnamefont {Y.}~\bibnamefont {Wu}}, \bibinfo {author} {\bibfnamefont {E.}~\bibnamefont {Timmons}}, \bibinfo {author} {\bibfnamefont {T.-H.}\ \bibnamefont {Kim}}, \bibinfo {author} {\bibfnamefont {L.}~\bibnamefont {Zhou}}, \bibinfo {author} {\bibfnamefont {L.-L.}\ \bibnamefont {Wang}}, \bibinfo {author} {\bibfnamefont {B.~G.}\ \bibnamefont {Ueland}}, \bibinfo {author} {\bibfnamefont {A.}~\bibnamefont {Palasyuk}}, \bibinfo {author} {\bibfnamefont {D.~H.}\ \bibnamefont {Ryan}}, \bibinfo {author} {\bibfnamefont {R.~J.}\ \bibnamefont {McQueeney}}, \bibinfo {author} {\bibfnamefont {K.}~\bibnamefont {Lee}}, \bibinfo {author} {\bibfnamefont {B.}~\bibnamefont {Schrunk}}, \bibinfo {author} {\bibfnamefont {A.~A.}\ \bibnamefont {Burkov}}, \bibinfo {author} {\bibfnamefont {R.}~\bibnamefont {Prozorov}}, \bibinfo {author} {\bibfnamefont {S.~L.}\
  \bibnamefont {Bud'ko}}, \bibinfo {author} {\bibfnamefont {A.}~\bibnamefont {Kaminski}},\ and\ \bibinfo {author} {\bibfnamefont {P.~C.}\ \bibnamefont {Canfield}},\ }\bibfield  {title} {\bibinfo {title} {Manipulating magnetism in the topological semimetal \ch{EuCd2As2}},\ }\href {https://doi.org/10.1103/PhysRevB.101.140402} {\bibfield  {journal} {\bibinfo  {journal} {Physical Review B}\ }\textbf {\bibinfo {volume} {101}},\ \bibinfo {pages} {140402} (\bibinfo {year} {2020})}\BibitemShut {NoStop}%
\bibitem [{\citenamefont {Xu}\ \emph {et~al.}(2019)\citenamefont {Xu}, \citenamefont {Song}, \citenamefont {Wang}, \citenamefont {Weng},\ and\ \citenamefont {Dai}}]{xu_higher-order_2019}%
  \BibitemOpen
  \bibfield  {author} {\bibinfo {author} {\bibfnamefont {Y.}~\bibnamefont {Xu}}, \bibinfo {author} {\bibfnamefont {Z.}~\bibnamefont {Song}}, \bibinfo {author} {\bibfnamefont {Z.}~\bibnamefont {Wang}}, \bibinfo {author} {\bibfnamefont {H.}~\bibnamefont {Weng}},\ and\ \bibinfo {author} {\bibfnamefont {X.}~\bibnamefont {Dai}},\ }\bibfield  {title} {\bibinfo {title} {Higher-{Order} {Topology} of the {Axion} {Insulator} \ch{EuIn2As2}},\ }\href {https://doi.org/10.1103/PhysRevLett.122.256402} {\bibfield  {journal} {\bibinfo  {journal} {Physical Review Letters}\ }\textbf {\bibinfo {volume} {122}},\ \bibinfo {pages} {256402} (\bibinfo {year} {2019})}\BibitemShut {NoStop}%
\bibitem [{\citenamefont {Riberolles}\ \emph {et~al.}(2021)\citenamefont {Riberolles}, \citenamefont {Trevisan}, \citenamefont {Kuthanazhi}, \citenamefont {Heitmann}, \citenamefont {Ye}, \citenamefont {Johnston}, \citenamefont {Bud’ko}, \citenamefont {Ryan}, \citenamefont {Canfield}, \citenamefont {Kreyssig}, \citenamefont {Vishwanath}, \citenamefont {McQueeney}, \citenamefont {Wang}, \citenamefont {Orth},\ and\ \citenamefont {Ueland}}]{riberolles_magnetic_2021}%
  \BibitemOpen
  \bibfield  {author} {\bibinfo {author} {\bibfnamefont {S.~X.~M.}\ \bibnamefont {Riberolles}}, \bibinfo {author} {\bibfnamefont {T.~V.}\ \bibnamefont {Trevisan}}, \bibinfo {author} {\bibfnamefont {B.}~\bibnamefont {Kuthanazhi}}, \bibinfo {author} {\bibfnamefont {T.~W.}\ \bibnamefont {Heitmann}}, \bibinfo {author} {\bibfnamefont {F.}~\bibnamefont {Ye}}, \bibinfo {author} {\bibfnamefont {D.~C.}\ \bibnamefont {Johnston}}, \bibinfo {author} {\bibfnamefont {S.~L.}\ \bibnamefont {Bud’ko}}, \bibinfo {author} {\bibfnamefont {D.~H.}\ \bibnamefont {Ryan}}, \bibinfo {author} {\bibfnamefont {P.~C.}\ \bibnamefont {Canfield}}, \bibinfo {author} {\bibfnamefont {A.}~\bibnamefont {Kreyssig}}, \bibinfo {author} {\bibfnamefont {A.}~\bibnamefont {Vishwanath}}, \bibinfo {author} {\bibfnamefont {R.~J.}\ \bibnamefont {McQueeney}}, \bibinfo {author} {\bibfnamefont {L.-L.}\ \bibnamefont {Wang}}, \bibinfo {author} {\bibfnamefont {P.~P.}\ \bibnamefont {Orth}},\ and\ \bibinfo {author} {\bibfnamefont {B.~G.}\ \bibnamefont {Ueland}},\
  }\bibfield  {title} {\bibinfo {title} {Magnetic crystalline-symmetry-protected axion electrodynamics and field-tunable unpinned {Dirac} cones in \ch{EuIn2As2}},\ }\href {https://doi.org/10.1038/s41467-021-21154-y} {\bibfield  {journal} {\bibinfo  {journal} {Nature Communications}\ }\textbf {\bibinfo {volume} {12}},\ \bibinfo {pages} {999} (\bibinfo {year} {2021})}\BibitemShut {NoStop}%
\bibitem [{\citenamefont {Wang}\ \emph {et~al.}(2021{\natexlab{a}})\citenamefont {Wang}, \citenamefont {Rogers}, \citenamefont {Yao}, \citenamefont {Nichols}, \citenamefont {Atay}, \citenamefont {Xu}, \citenamefont {Franklin}, \citenamefont {Sochnikov}, \citenamefont {Ryan}, \citenamefont {Haskel},\ and\ \citenamefont {Tafti}}]{wang_colossal_2021}%
  \BibitemOpen
  \bibfield  {author} {\bibinfo {author} {\bibfnamefont {Z.-C.}\ \bibnamefont {Wang}}, \bibinfo {author} {\bibfnamefont {J.~D.}\ \bibnamefont {Rogers}}, \bibinfo {author} {\bibfnamefont {X.}~\bibnamefont {Yao}}, \bibinfo {author} {\bibfnamefont {R.}~\bibnamefont {Nichols}}, \bibinfo {author} {\bibfnamefont {K.}~\bibnamefont {Atay}}, \bibinfo {author} {\bibfnamefont {B.}~\bibnamefont {Xu}}, \bibinfo {author} {\bibfnamefont {J.}~\bibnamefont {Franklin}}, \bibinfo {author} {\bibfnamefont {I.}~\bibnamefont {Sochnikov}}, \bibinfo {author} {\bibfnamefont {P.~J.}\ \bibnamefont {Ryan}}, \bibinfo {author} {\bibfnamefont {D.}~\bibnamefont {Haskel}},\ and\ \bibinfo {author} {\bibfnamefont {F.}~\bibnamefont {Tafti}},\ }\bibfield  {title} {\bibinfo {title} {Colossal {Magnetoresistance} without {Mixed} {Valence} in a {Layered} {Phosphide} {Crystal}},\ }\href {https://doi.org/10.1002/adma.202005755} {\bibfield  {journal} {\bibinfo  {journal} {Advanced Materials}\ }\textbf {\bibinfo {volume} {33}},\ \bibinfo {pages}
  {2005755} (\bibinfo {year} {2021}{\natexlab{a}})}\BibitemShut {NoStop}%
\bibitem [{\citenamefont {Sunko}\ \emph {et~al.}(2023)\citenamefont {Sunko}, \citenamefont {Sun}, \citenamefont {Vranas}, \citenamefont {Homes}, \citenamefont {Lee}, \citenamefont {Donoway}, \citenamefont {Wang}, \citenamefont {Balguri}, \citenamefont {Mahendru}, \citenamefont {Ruiz}, \citenamefont {Gunn}, \citenamefont {Basak}, \citenamefont {Blanco-Canosa}, \citenamefont {Schierle}, \citenamefont {Weschke}, \citenamefont {Tafti}, \citenamefont {Frano},\ and\ \citenamefont {Orenstein}}]{sunko_spin-carrier_2023}%
  \BibitemOpen
  \bibfield  {author} {\bibinfo {author} {\bibfnamefont {V.}~\bibnamefont {Sunko}}, \bibinfo {author} {\bibfnamefont {Y.}~\bibnamefont {Sun}}, \bibinfo {author} {\bibfnamefont {M.}~\bibnamefont {Vranas}}, \bibinfo {author} {\bibfnamefont {C.~C.}\ \bibnamefont {Homes}}, \bibinfo {author} {\bibfnamefont {C.}~\bibnamefont {Lee}}, \bibinfo {author} {\bibfnamefont {E.}~\bibnamefont {Donoway}}, \bibinfo {author} {\bibfnamefont {Z.-C.}\ \bibnamefont {Wang}}, \bibinfo {author} {\bibfnamefont {S.}~\bibnamefont {Balguri}}, \bibinfo {author} {\bibfnamefont {M.~B.}\ \bibnamefont {Mahendru}}, \bibinfo {author} {\bibfnamefont {A.}~\bibnamefont {Ruiz}}, \bibinfo {author} {\bibfnamefont {B.}~\bibnamefont {Gunn}}, \bibinfo {author} {\bibfnamefont {R.}~\bibnamefont {Basak}}, \bibinfo {author} {\bibfnamefont {S.}~\bibnamefont {Blanco-Canosa}}, \bibinfo {author} {\bibfnamefont {E.}~\bibnamefont {Schierle}}, \bibinfo {author} {\bibfnamefont {E.}~\bibnamefont {Weschke}}, \bibinfo {author} {\bibfnamefont {F.}~\bibnamefont {Tafti}},
  \bibinfo {author} {\bibfnamefont {A.}~\bibnamefont {Frano}},\ and\ \bibinfo {author} {\bibfnamefont {J.}~\bibnamefont {Orenstein}},\ }\bibfield  {title} {\bibinfo {title} {Spin-carrier coupling induced ferromagnetism and giant resistivity peak in \ch{EuCd2P2}},\ }\href {https://doi.org/10.1103/PhysRevB.107.144404} {\bibfield  {journal} {\bibinfo  {journal} {Physical Review B}\ }\textbf {\bibinfo {volume} {107}},\ \bibinfo {pages} {144404} (\bibinfo {year} {2023})}\BibitemShut {NoStop}%
\bibitem [{\citenamefont {Homes}\ \emph {et~al.}(2023)\citenamefont {Homes}, \citenamefont {Wang}, \citenamefont {Fruhling},\ and\ \citenamefont {Tafti}}]{homes_optical_2023}%
  \BibitemOpen
  \bibfield  {author} {\bibinfo {author} {\bibfnamefont {C.~C.}\ \bibnamefont {Homes}}, \bibinfo {author} {\bibfnamefont {Z.-C.}\ \bibnamefont {Wang}}, \bibinfo {author} {\bibfnamefont {K.}~\bibnamefont {Fruhling}},\ and\ \bibinfo {author} {\bibfnamefont {F.}~\bibnamefont {Tafti}},\ }\bibfield  {title} {\bibinfo {title} {Optical properties and carrier localization in the layered phosphide \ch{EuCd2P2}},\ }\href {https://doi.org/10.1103/PhysRevB.107.045106} {\bibfield  {journal} {\bibinfo  {journal} {Physical Review B}\ }\textbf {\bibinfo {volume} {107}},\ \bibinfo {pages} {045106} (\bibinfo {year} {2023})}\BibitemShut {NoStop}%
\bibitem [{\citenamefont {Rosa}\ \emph {et~al.}(2020)\citenamefont {Rosa}, \citenamefont {Xu}, \citenamefont {Rahn}, \citenamefont {Souza}, \citenamefont {Kushwaha}, \citenamefont {Veiga}, \citenamefont {Bombardi}, \citenamefont {Thomas}, \citenamefont {Janoschek}, \citenamefont {Bauer}, \citenamefont {Chan}, \citenamefont {Wang}, \citenamefont {Thompson}, \citenamefont {Harrison}, \citenamefont {Pagliuso}, \citenamefont {Bernevig},\ and\ \citenamefont {Ronning}}]{rosa_colossal_2020}%
  \BibitemOpen
  \bibfield  {author} {\bibinfo {author} {\bibfnamefont {P.}~\bibnamefont {Rosa}}, \bibinfo {author} {\bibfnamefont {Y.}~\bibnamefont {Xu}}, \bibinfo {author} {\bibfnamefont {M.}~\bibnamefont {Rahn}}, \bibinfo {author} {\bibfnamefont {J.}~\bibnamefont {Souza}}, \bibinfo {author} {\bibfnamefont {S.}~\bibnamefont {Kushwaha}}, \bibinfo {author} {\bibfnamefont {L.}~\bibnamefont {Veiga}}, \bibinfo {author} {\bibfnamefont {A.}~\bibnamefont {Bombardi}}, \bibinfo {author} {\bibfnamefont {S.}~\bibnamefont {Thomas}}, \bibinfo {author} {\bibfnamefont {M.}~\bibnamefont {Janoschek}}, \bibinfo {author} {\bibfnamefont {E.}~\bibnamefont {Bauer}}, \bibinfo {author} {\bibfnamefont {M.}~\bibnamefont {Chan}}, \bibinfo {author} {\bibfnamefont {Z.}~\bibnamefont {Wang}}, \bibinfo {author} {\bibfnamefont {J.}~\bibnamefont {Thompson}}, \bibinfo {author} {\bibfnamefont {N.}~\bibnamefont {Harrison}}, \bibinfo {author} {\bibfnamefont {P.}~\bibnamefont {Pagliuso}}, \bibinfo {author} {\bibfnamefont {A.}~\bibnamefont {Bernevig}},\ and\
  \bibinfo {author} {\bibfnamefont {F.}~\bibnamefont {Ronning}},\ }\bibfield  {title} {\bibinfo {title} {Colossal magnetoresistance in a nonsymmorphic antiferromagnetic insulator},\ }\href {https://doi.org/10.1038/s41535-020-00256-8} {\bibfield  {journal} {\bibinfo  {journal} {npj Quantum Materials}\ }\textbf {\bibinfo {volume} {5}},\ \bibinfo {pages} {1} (\bibinfo {year} {2020})}\BibitemShut {NoStop}%
\bibitem [{\citenamefont {Ghosh}\ \emph {et~al.}(2022)\citenamefont {Ghosh}, \citenamefont {Lane}, \citenamefont {Ronning}, \citenamefont {Bauer}, \citenamefont {Thompson}, \citenamefont {Zhu}, \citenamefont {Rosa},\ and\ \citenamefont {Thomas}}]{ghosh_colossal_2022}%
  \BibitemOpen
  \bibfield  {author} {\bibinfo {author} {\bibfnamefont {S.}~\bibnamefont {Ghosh}}, \bibinfo {author} {\bibfnamefont {C.}~\bibnamefont {Lane}}, \bibinfo {author} {\bibfnamefont {F.}~\bibnamefont {Ronning}}, \bibinfo {author} {\bibfnamefont {E.~D.}\ \bibnamefont {Bauer}}, \bibinfo {author} {\bibfnamefont {J.~D.}\ \bibnamefont {Thompson}}, \bibinfo {author} {\bibfnamefont {J.-X.}\ \bibnamefont {Zhu}}, \bibinfo {author} {\bibfnamefont {P.~F.~S.}\ \bibnamefont {Rosa}},\ and\ \bibinfo {author} {\bibfnamefont {S.~M.}\ \bibnamefont {Thomas}},\ }\bibfield  {title} {\bibinfo {title} {Colossal piezoresistance in narrow-gap \ch{Eu5In2Sb6}},\ }\href {https://doi.org/10.1103/PhysRevB.106.045110} {\bibfield  {journal} {\bibinfo  {journal} {Physical Review B}\ }\textbf {\bibinfo {volume} {106}},\ \bibinfo {pages} {045110} (\bibinfo {year} {2022})}\BibitemShut {NoStop}%
\bibitem [{\citenamefont {Morano}\ \emph {et~al.}(2024)\citenamefont {Morano}, \citenamefont {Gaudet}, \citenamefont {Varnava}, \citenamefont {Berry}, \citenamefont {Halloran}, \citenamefont {Lygouras}, \citenamefont {Wang}, \citenamefont {Hoffman}, \citenamefont {Xu}, \citenamefont {Lynn}, \citenamefont {McQueen}, \citenamefont {Vanderbilt},\ and\ \citenamefont {Broholm}}]{morano_noncollinear_2024}%
  \BibitemOpen
  \bibfield  {author} {\bibinfo {author} {\bibfnamefont {V.~C.}\ \bibnamefont {Morano}}, \bibinfo {author} {\bibfnamefont {J.}~\bibnamefont {Gaudet}}, \bibinfo {author} {\bibfnamefont {N.}~\bibnamefont {Varnava}}, \bibinfo {author} {\bibfnamefont {T.}~\bibnamefont {Berry}}, \bibinfo {author} {\bibfnamefont {T.}~\bibnamefont {Halloran}}, \bibinfo {author} {\bibfnamefont {C.~J.}\ \bibnamefont {Lygouras}}, \bibinfo {author} {\bibfnamefont {X.}~\bibnamefont {Wang}}, \bibinfo {author} {\bibfnamefont {C.~M.}\ \bibnamefont {Hoffman}}, \bibinfo {author} {\bibfnamefont {G.}~\bibnamefont {Xu}}, \bibinfo {author} {\bibfnamefont {J.~W.}\ \bibnamefont {Lynn}}, \bibinfo {author} {\bibfnamefont {T.~M.}\ \bibnamefont {McQueen}}, \bibinfo {author} {\bibfnamefont {D.}~\bibnamefont {Vanderbilt}},\ and\ \bibinfo {author} {\bibfnamefont {C.~L.}\ \bibnamefont {Broholm}},\ }\bibfield  {title} {\bibinfo {title} {Noncollinear $2k$ antiferromagnetism in the {Zintl} semiconductor \ch{Eu5In2Sb6}},\ }\href
  {https://doi.org/10.1103/PhysRevB.109.014432} {\bibfield  {journal} {\bibinfo  {journal} {Physical Review B}\ }\textbf {\bibinfo {volume} {109}},\ \bibinfo {pages} {014432} (\bibinfo {year} {2024})}\BibitemShut {NoStop}%
\bibitem [{\citenamefont {Souza}\ \emph {et~al.}(2022)\citenamefont {Souza}, \citenamefont {Thomas}, \citenamefont {Bauer}, \citenamefont {Thompson}, \citenamefont {Ronning}, \citenamefont {Pagliuso},\ and\ \citenamefont {Rosa}}]{souza_microscopic_2022}%
  \BibitemOpen
  \bibfield  {author} {\bibinfo {author} {\bibfnamefont {J.~C.}\ \bibnamefont {Souza}}, \bibinfo {author} {\bibfnamefont {S.~M.}\ \bibnamefont {Thomas}}, \bibinfo {author} {\bibfnamefont {E.~D.}\ \bibnamefont {Bauer}}, \bibinfo {author} {\bibfnamefont {J.~D.}\ \bibnamefont {Thompson}}, \bibinfo {author} {\bibfnamefont {F.}~\bibnamefont {Ronning}}, \bibinfo {author} {\bibfnamefont {P.~G.}\ \bibnamefont {Pagliuso}},\ and\ \bibinfo {author} {\bibfnamefont {P.~F.~S.}\ \bibnamefont {Rosa}},\ }\bibfield  {title} {\bibinfo {title} {Microscopic probe of magnetic polarons in antiferromagnetic \ch{Eu5In2As6}},\ }\href {https://doi.org/10.1103/PhysRevB.105.035135} {\bibfield  {journal} {\bibinfo  {journal} {Physical Review B}\ }\textbf {\bibinfo {volume} {105}},\ \bibinfo {pages} {035135} (\bibinfo {year} {2022})}\BibitemShut {NoStop}%
\bibitem [{\citenamefont {Rahn}\ \emph {et~al.}(2023)\citenamefont {Rahn}, \citenamefont {Wilson}, \citenamefont {Hicken}, \citenamefont {Pratt}, \citenamefont {Wang}, \citenamefont {Orlandi}, \citenamefont {Khalyavin}, \citenamefont {Manuel}, \citenamefont {Veiga}, \citenamefont {Bombardi}, \citenamefont {Francoual}, \citenamefont {Bereciartua}, \citenamefont {Sukhanov}, \citenamefont {Thompson}, \citenamefont {Thomas}, \citenamefont {Rosa}, \citenamefont {Lancaster}, \citenamefont {Ronning},\ and\ \citenamefont {Janoschek}}]{rahn_unusual_2023}%
  \BibitemOpen
  \bibfield  {author} {\bibinfo {author} {\bibfnamefont {M.~C.}\ \bibnamefont {Rahn}}, \bibinfo {author} {\bibfnamefont {M.~N.}\ \bibnamefont {Wilson}}, \bibinfo {author} {\bibfnamefont {T.~J.}\ \bibnamefont {Hicken}}, \bibinfo {author} {\bibfnamefont {F.~L.}\ \bibnamefont {Pratt}}, \bibinfo {author} {\bibfnamefont {C.}~\bibnamefont {Wang}}, \bibinfo {author} {\bibfnamefont {F.}~\bibnamefont {Orlandi}}, \bibinfo {author} {\bibfnamefont {D.~D.}\ \bibnamefont {Khalyavin}}, \bibinfo {author} {\bibfnamefont {P.}~\bibnamefont {Manuel}}, \bibinfo {author} {\bibfnamefont {L.~S.~I.}\ \bibnamefont {Veiga}}, \bibinfo {author} {\bibfnamefont {A.}~\bibnamefont {Bombardi}}, \bibinfo {author} {\bibfnamefont {S.}~\bibnamefont {Francoual}}, \bibinfo {author} {\bibfnamefont {P.}~\bibnamefont {Bereciartua}}, \bibinfo {author} {\bibfnamefont {A.~S.}\ \bibnamefont {Sukhanov}}, \bibinfo {author} {\bibfnamefont {J.~D.}\ \bibnamefont {Thompson}}, \bibinfo {author} {\bibfnamefont {S.~M.}\ \bibnamefont {Thomas}}, \bibinfo {author}
  {\bibfnamefont {P.~F.~S.}\ \bibnamefont {Rosa}}, \bibinfo {author} {\bibfnamefont {T.}~\bibnamefont {Lancaster}}, \bibinfo {author} {\bibfnamefont {F.}~\bibnamefont {Ronning}},\ and\ \bibinfo {author} {\bibfnamefont {M.}~\bibnamefont {Janoschek}},\ }\href {https://doi.org/10.48550/arXiv.2312.15054} {\bibinfo {title} {Unusual magnetism of the axion-insulator candidate \ch{Eu5In2Sb6}}} (\bibinfo {year} {2023})\BibitemShut {NoStop}%
\bibitem [{\citenamefont {Zhang}\ \emph {et~al.}(2022)\citenamefont {Zhang}, \citenamefont {Ni}, \citenamefont {Zhao}, \citenamefont {Hakani}, \citenamefont {Ye}, \citenamefont {DeLong}, \citenamefont {Kimchi},\ and\ \citenamefont {Cao}}]{zhang_control_2022}%
  \BibitemOpen
  \bibfield  {author} {\bibinfo {author} {\bibfnamefont {Y.}~\bibnamefont {Zhang}}, \bibinfo {author} {\bibfnamefont {Y.}~\bibnamefont {Ni}}, \bibinfo {author} {\bibfnamefont {H.}~\bibnamefont {Zhao}}, \bibinfo {author} {\bibfnamefont {S.}~\bibnamefont {Hakani}}, \bibinfo {author} {\bibfnamefont {F.}~\bibnamefont {Ye}}, \bibinfo {author} {\bibfnamefont {L.}~\bibnamefont {DeLong}}, \bibinfo {author} {\bibfnamefont {I.}~\bibnamefont {Kimchi}},\ and\ \bibinfo {author} {\bibfnamefont {G.}~\bibnamefont {Cao}},\ }\bibfield  {title} {\bibinfo {title} {Control of chiral orbital currents in a colossal magnetoresistance material},\ }\href {https://doi.org/10.1038/s41586-022-05262-3} {\bibfield  {journal} {\bibinfo  {journal} {Nature}\ }\textbf {\bibinfo {volume} {611}},\ \bibinfo {pages} {467} (\bibinfo {year} {2022})}\BibitemShut {NoStop}%
\bibitem [{\citenamefont {Singha}\ \emph {et~al.}(2023)\citenamefont {Singha}, \citenamefont {Dalgaard}, \citenamefont {Marchenko}, \citenamefont {Krivenkov}, \citenamefont {Rienks}, \citenamefont {Jovanovic}, \citenamefont {Teicher}, \citenamefont {Hu}, \citenamefont {Salters}, \citenamefont {Lin}, \citenamefont {Varykhalov}, \citenamefont {Ong},\ and\ \citenamefont {Schoop}}]{singha_colossal_2023}%
  \BibitemOpen
  \bibfield  {author} {\bibinfo {author} {\bibfnamefont {R.}~\bibnamefont {Singha}}, \bibinfo {author} {\bibfnamefont {K.~J.}\ \bibnamefont {Dalgaard}}, \bibinfo {author} {\bibfnamefont {D.}~\bibnamefont {Marchenko}}, \bibinfo {author} {\bibfnamefont {M.}~\bibnamefont {Krivenkov}}, \bibinfo {author} {\bibfnamefont {E.~D.~L.}\ \bibnamefont {Rienks}}, \bibinfo {author} {\bibfnamefont {M.}~\bibnamefont {Jovanovic}}, \bibinfo {author} {\bibfnamefont {S.~M.~L.}\ \bibnamefont {Teicher}}, \bibinfo {author} {\bibfnamefont {J.}~\bibnamefont {Hu}}, \bibinfo {author} {\bibfnamefont {T.~H.}\ \bibnamefont {Salters}}, \bibinfo {author} {\bibfnamefont {J.}~\bibnamefont {Lin}}, \bibinfo {author} {\bibfnamefont {A.}~\bibnamefont {Varykhalov}}, \bibinfo {author} {\bibfnamefont {N.~P.}\ \bibnamefont {Ong}},\ and\ \bibinfo {author} {\bibfnamefont {L.~M.}\ \bibnamefont {Schoop}},\ }\bibfield  {title} {\bibinfo {title} {Colossal magnetoresistance in the multiple wave vector charge density wave regime of an antiferromagnetic
  {Dirac} semimetal},\ }\href {https://doi.org/10.1126/sciadv.adh0145} {\bibfield  {journal} {\bibinfo  {journal} {Science Advances}\ }\textbf {\bibinfo {volume} {9}},\ \bibinfo {pages} {eadh0145} (\bibinfo {year} {2023})}\BibitemShut {NoStop}%
\bibitem [{\citenamefont {Ram}\ \emph {et~al.}(2024)\citenamefont {Ram}, \citenamefont {Singh}, \citenamefont {Banerjee}, \citenamefont {Sundaresan}, \citenamefont {Samal}, \citenamefont {Kanchana},\ and\ \citenamefont {Hossain}}]{ram_magnetotransport_2024}%
  \BibitemOpen
  \bibfield  {author} {\bibinfo {author} {\bibfnamefont {D.}~\bibnamefont {Ram}}, \bibinfo {author} {\bibfnamefont {J.}~\bibnamefont {Singh}}, \bibinfo {author} {\bibfnamefont {S.}~\bibnamefont {Banerjee}}, \bibinfo {author} {\bibfnamefont {A.}~\bibnamefont {Sundaresan}}, \bibinfo {author} {\bibfnamefont {D.}~\bibnamefont {Samal}}, \bibinfo {author} {\bibfnamefont {V.}~\bibnamefont {Kanchana}},\ and\ \bibinfo {author} {\bibfnamefont {Z.}~\bibnamefont {Hossain}},\ }\bibfield  {title} {\bibinfo {title} {Magnetotransport and electronic structure of {EuAuSb}: {A} candidate antiferromagnetic {Dirac} semimetal},\ }\href {https://doi.org/10.1103/PhysRevB.109.155152} {\bibfield  {journal} {\bibinfo  {journal} {Physical Review B}\ }\textbf {\bibinfo {volume} {109}},\ \bibinfo {pages} {155152} (\bibinfo {year} {2024})}\BibitemShut {NoStop}%
\bibitem [{\citenamefont {Zhang}\ \emph {et~al.}(2023)\citenamefont {Zhang}, \citenamefont {Du}, \citenamefont {Zheng}, \citenamefont {Luo}, \citenamefont {Wu}, \citenamefont {Zheng}, \citenamefont {Cui}, \citenamefont {Sun}, \citenamefont {Liu}, \citenamefont {Shen}, \citenamefont {Smidman}, \citenamefont {Song}, \citenamefont {Shi}, \citenamefont {Zhong}, \citenamefont {Cao}, \citenamefont {Yuan},\ and\ \citenamefont {Liu}}]{zhang_electronic_2023}%
  \BibitemOpen
  \bibfield  {author} {\bibinfo {author} {\bibfnamefont {H.}~\bibnamefont {Zhang}}, \bibinfo {author} {\bibfnamefont {F.}~\bibnamefont {Du}}, \bibinfo {author} {\bibfnamefont {X.}~\bibnamefont {Zheng}}, \bibinfo {author} {\bibfnamefont {S.}~\bibnamefont {Luo}}, \bibinfo {author} {\bibfnamefont {Y.}~\bibnamefont {Wu}}, \bibinfo {author} {\bibfnamefont {H.}~\bibnamefont {Zheng}}, \bibinfo {author} {\bibfnamefont {S.}~\bibnamefont {Cui}}, \bibinfo {author} {\bibfnamefont {Z.}~\bibnamefont {Sun}}, \bibinfo {author} {\bibfnamefont {Z.}~\bibnamefont {Liu}}, \bibinfo {author} {\bibfnamefont {D.}~\bibnamefont {Shen}}, \bibinfo {author} {\bibfnamefont {M.}~\bibnamefont {Smidman}}, \bibinfo {author} {\bibfnamefont {Y.}~\bibnamefont {Song}}, \bibinfo {author} {\bibfnamefont {M.}~\bibnamefont {Shi}}, \bibinfo {author} {\bibfnamefont {Z.}~\bibnamefont {Zhong}}, \bibinfo {author} {\bibfnamefont {C.}~\bibnamefont {Cao}}, \bibinfo {author} {\bibfnamefont {H.}~\bibnamefont {Yuan}},\ and\ \bibinfo {author} {\bibfnamefont
  {Y.}~\bibnamefont {Liu}},\ }\bibfield  {title} {\bibinfo {title} {Electronic band reconstruction across the insulator-metal transition in colossally magnetoresistive \ch{EuCd2P2}},\ }\href {https://doi.org/10.1103/PhysRevB.108.L241115} {\bibfield  {journal} {\bibinfo  {journal} {Physical Review B}\ }\textbf {\bibinfo {volume} {108}},\ \bibinfo {pages} {L241115} (\bibinfo {year} {2023})}\BibitemShut {NoStop}%
\bibitem [{\citenamefont {Varnava}\ \emph {et~al.}(2022)\citenamefont {Varnava}, \citenamefont {Berry}, \citenamefont {McQueen},\ and\ \citenamefont {Vanderbilt}}]{varnava_engineering_2022}%
  \BibitemOpen
  \bibfield  {author} {\bibinfo {author} {\bibfnamefont {N.}~\bibnamefont {Varnava}}, \bibinfo {author} {\bibfnamefont {T.}~\bibnamefont {Berry}}, \bibinfo {author} {\bibfnamefont {T.~M.}\ \bibnamefont {McQueen}},\ and\ \bibinfo {author} {\bibfnamefont {D.}~\bibnamefont {Vanderbilt}},\ }\bibfield  {title} {\bibinfo {title} {Engineering magnetic topological insulators in \ch{Eu5M2X6} {Zintl} compounds},\ }\href {https://doi.org/10.1103/PhysRevB.105.235128} {\bibfield  {journal} {\bibinfo  {journal} {Physical Review B}\ }\textbf {\bibinfo {volume} {105}},\ \bibinfo {pages} {235128} (\bibinfo {year} {2022})}\BibitemShut {NoStop}%
\bibitem [{\citenamefont {Stern}(1958)}]{stern_theory_1958}%
  \BibitemOpen
  \bibfield  {author} {\bibinfo {author} {\bibfnamefont {E.~A.}\ \bibnamefont {Stern}},\ }\bibfield  {title} {\bibinfo {title} {Theory of the {Anharmonic} {Properties} of {Solids}},\ }\href {https://doi.org/10.1103/PhysRev.111.786} {\bibfield  {journal} {\bibinfo  {journal} {Physical Review}\ }\textbf {\bibinfo {volume} {111}},\ \bibinfo {pages} {786} (\bibinfo {year} {1958})}\BibitemShut {NoStop}%
\bibitem [{sup()}]{suppmatt}%
  \BibitemOpen
  \href {https://journals.aps.org} {}\bibinfo {note} {See the Supplemental Material for additional information about crystal growth, Curie-Weiss analysis, low-temperature specific heat, tunnel diode oscillations, and transport data.}\BibitemShut {Stop}%
\bibitem [{\citenamefont {Tomitaka}\ \emph {et~al.}(2021)\citenamefont {Tomitaka}, \citenamefont {Goto}, \citenamefont {Morino}, \citenamefont {Hoshi}, \citenamefont {Nakahira}, \citenamefont {Ito}, \citenamefont {Miura}, \citenamefont {Usui},\ and\ \citenamefont {Mizuguchi}}]{tomitaka_bipolar_2021}%
  \BibitemOpen
  \bibfield  {author} {\bibinfo {author} {\bibfnamefont {N.}~\bibnamefont {Tomitaka}}, \bibinfo {author} {\bibfnamefont {Y.}~\bibnamefont {Goto}}, \bibinfo {author} {\bibfnamefont {K.}~\bibnamefont {Morino}}, \bibinfo {author} {\bibfnamefont {K.}~\bibnamefont {Hoshi}}, \bibinfo {author} {\bibfnamefont {Y.}~\bibnamefont {Nakahira}}, \bibinfo {author} {\bibfnamefont {H.}~\bibnamefont {Ito}}, \bibinfo {author} {\bibfnamefont {A.}~\bibnamefont {Miura}}, \bibinfo {author} {\bibfnamefont {H.}~\bibnamefont {Usui}},\ and\ \bibinfo {author} {\bibfnamefont {Y.}~\bibnamefont {Mizuguchi}},\ }\bibfield  {title} {\bibinfo {title} {Bipolar doping and thermoelectric properties of {Zintl} arsenide \ch{Eu5In2As6}},\ }\href {https://doi.org/10.1039/D1TA07559D} {\bibfield  {journal} {\bibinfo  {journal} {Journal of Materials Chemistry A}\ }\textbf {\bibinfo {volume} {9}},\ \bibinfo {pages} {26362} (\bibinfo {year} {2021})}\BibitemShut {NoStop}%
\bibitem [{\citenamefont {Radzieowski}\ \emph {et~al.}(2020)\citenamefont {Radzieowski}, \citenamefont {Stegemann}, \citenamefont {Klenner}, \citenamefont {Zhang}, \citenamefont {Fokwa},\ and\ \citenamefont {Janka}}]{radzieowski_divalent_2020}%
  \BibitemOpen
  \bibfield  {author} {\bibinfo {author} {\bibfnamefont {M.}~\bibnamefont {Radzieowski}}, \bibinfo {author} {\bibfnamefont {F.}~\bibnamefont {Stegemann}}, \bibinfo {author} {\bibfnamefont {S.}~\bibnamefont {Klenner}}, \bibinfo {author} {\bibfnamefont {Y.}~\bibnamefont {Zhang}}, \bibinfo {author} {\bibfnamefont {B.~P.~T.}\ \bibnamefont {Fokwa}},\ and\ \bibinfo {author} {\bibfnamefont {O.}~\bibnamefont {Janka}},\ }\bibfield  {title} {\bibinfo {title} {On the divalent character of the {Eu} atoms in the ternary {Zintl} phases \ch{Eu5In2Pn6} and \ch{Eu3MAs3} ({Pn} = {As}–{Bi}; {M} = {Al}, {Ga})},\ }\href {https://doi.org/10.1039/C9QM00703B} {\bibfield  {journal} {\bibinfo  {journal} {Materials Chemistry Frontiers}\ }\textbf {\bibinfo {volume} {4}},\ \bibinfo {pages} {1231} (\bibinfo {year} {2020})}\BibitemShut {NoStop}%
\bibitem [{\citenamefont {Rodríguez-Carvajal}(1993)}]{rodriguez-carvajal_recent_1993}%
  \BibitemOpen
  \bibfield  {author} {\bibinfo {author} {\bibfnamefont {J.}~\bibnamefont {Rodríguez-Carvajal}},\ }\bibfield  {title} {\bibinfo {title} {Recent advances in magnetic structure determination by neutron powder diffraction},\ }\href {https://doi.org/10.1016/0921-4526(93)90108-I} {\bibfield  {journal} {\bibinfo  {journal} {Physica B: Condensed Matter}\ }\textbf {\bibinfo {volume} {192}},\ \bibinfo {pages} {55} (\bibinfo {year} {1993})}\BibitemShut {NoStop}%
\bibitem [{\citenamefont {Momma}\ and\ \citenamefont {Izumi}(2011)}]{momma_vesta_2011}%
  \BibitemOpen
  \bibfield  {author} {\bibinfo {author} {\bibfnamefont {K.}~\bibnamefont {Momma}}\ and\ \bibinfo {author} {\bibfnamefont {F.}~\bibnamefont {Izumi}},\ }\bibfield  {title} {\bibinfo {title} {{VESTA} 3 for three-dimensional visualization of crystal, volumetric and morphology data},\ }\href {https://doi.org/10.1107/S0021889811038970} {\bibfield  {journal} {\bibinfo  {journal} {Journal of Applied Crystallography}\ }\textbf {\bibinfo {volume} {44}},\ \bibinfo {pages} {1272} (\bibinfo {year} {2011})}\BibitemShut {NoStop}%
\bibitem [{\citenamefont {Wills}(2000)}]{wills_new_2000}%
  \BibitemOpen
  \bibfield  {author} {\bibinfo {author} {\bibfnamefont {A.~S.}\ \bibnamefont {Wills}},\ }\bibfield  {title} {\bibinfo {title} {A new protocol for the determination of magnetic structures using simulated annealing and representational analysis ({SARA}\textit{h})},\ }\href {https://doi.org/10.1016/S0921-4526(99)01722-6} {\bibfield  {journal} {\bibinfo  {journal} {Physica B: Condensed Matter}\ }\textbf {\bibinfo {volume} {276-278}},\ \bibinfo {pages} {680} (\bibinfo {year} {2000})}\BibitemShut {NoStop}%
\bibitem [{\citenamefont {Chesser}\ and\ \citenamefont {Axe}(1973)}]{chesser_derivation_1973}%
  \BibitemOpen
  \bibfield  {author} {\bibinfo {author} {\bibfnamefont {N.~J.}\ \bibnamefont {Chesser}}\ and\ \bibinfo {author} {\bibfnamefont {J.~D.}\ \bibnamefont {Axe}},\ }\bibfield  {title} {\bibinfo {title} {Derivation and experimental verification of the normalized resolution function for inelastic neutron scattering},\ }\href {https://doi.org/10.1107/S0567739473000422} {\bibfield  {journal} {\bibinfo  {journal} {Acta Crystallographica Section A: Crystal Physics, Diffraction, Theoretical and General Crystallography}\ }\textbf {\bibinfo {volume} {29}},\ \bibinfo {pages} {160} (\bibinfo {year} {1973})}\BibitemShut {NoStop}%
\bibitem [{\citenamefont {Wuensch}\ and\ \citenamefont {Prewitt}(1965)}]{wuensch_corrections_1965}%
  \BibitemOpen
  \bibfield  {author} {\bibinfo {author} {\bibfnamefont {B.~J.}\ \bibnamefont {Wuensch}}\ and\ \bibinfo {author} {\bibfnamefont {C.~T.}\ \bibnamefont {Prewitt}},\ }\bibfield  {title} {\bibinfo {title} {Corrections for x-ray absorption by a crystal of arbitrary shape},\ }\href {https://doi.org/10.1515/zkri-1965-1-603} {\bibfield  {journal} {\bibinfo  {journal} {Zeitschrift für Kristallographie - Crystalline Materials}\ }\textbf {\bibinfo {volume} {122}},\ \bibinfo {pages} {24} (\bibinfo {year} {1965})}\BibitemShut {NoStop}%
\bibitem [{\citenamefont {Kauzlarich}(2023)}]{kauzlarich_zintl_2023}%
  \BibitemOpen
  \bibfield  {author} {\bibinfo {author} {\bibfnamefont {S.~M.}\ \bibnamefont {Kauzlarich}},\ }\bibfield  {title} {\bibinfo {title} {Zintl {Phases}: {From} {Curiosities} to {Impactful} {Materials}},\ }\href {https://doi.org/10.1021/acs.chemmater.3c01874} {\bibfield  {journal} {\bibinfo  {journal} {Chemistry of Materials}\ }\textbf {\bibinfo {volume} {35}},\ \bibinfo {pages} {7355} (\bibinfo {year} {2023})}\BibitemShut {NoStop}%
\bibitem [{\citenamefont {Childs}\ \emph {et~al.}(2019)\citenamefont {Childs}, \citenamefont {Baranets},\ and\ \citenamefont {Bobev}}]{childs_five_2019}%
  \BibitemOpen
  \bibfield  {author} {\bibinfo {author} {\bibfnamefont {A.~B.}\ \bibnamefont {Childs}}, \bibinfo {author} {\bibfnamefont {S.}~\bibnamefont {Baranets}},\ and\ \bibinfo {author} {\bibfnamefont {S.}~\bibnamefont {Bobev}},\ }\bibfield  {title} {\bibinfo {title} {Five new ternary indium-arsenides discovered. {Synthesis} and structural characterization of the {Zintl} phases \ch{Sr3In2As4}, \ch{Ba3In2As4}, \ch{Eu3In2As4}, \ch{Sr5In2As6} and \ch{Eu5In2As6}},\ }\href {https://doi.org/10.1016/j.jssc.2019.07.050} {\bibfield  {journal} {\bibinfo  {journal} {Journal of Solid State Chemistry}\ }\textbf {\bibinfo {volume} {278}},\ \bibinfo {pages} {120889} (\bibinfo {year} {2019})}\BibitemShut {NoStop}%
\bibitem [{\citenamefont {Wieder}\ \emph {et~al.}(2018)\citenamefont {Wieder}, \citenamefont {Bradlyn}, \citenamefont {Wang}, \citenamefont {Cano}, \citenamefont {Kim}, \citenamefont {Kim}, \citenamefont {Rappe}, \citenamefont {Kane},\ and\ \citenamefont {Bernevig}}]{wieder_wallpaper_2018}%
  \BibitemOpen
  \bibfield  {author} {\bibinfo {author} {\bibfnamefont {B.~J.}\ \bibnamefont {Wieder}}, \bibinfo {author} {\bibfnamefont {B.}~\bibnamefont {Bradlyn}}, \bibinfo {author} {\bibfnamefont {Z.}~\bibnamefont {Wang}}, \bibinfo {author} {\bibfnamefont {J.}~\bibnamefont {Cano}}, \bibinfo {author} {\bibfnamefont {Y.}~\bibnamefont {Kim}}, \bibinfo {author} {\bibfnamefont {H.-S.~D.}\ \bibnamefont {Kim}}, \bibinfo {author} {\bibfnamefont {A.~M.}\ \bibnamefont {Rappe}}, \bibinfo {author} {\bibfnamefont {C.~L.}\ \bibnamefont {Kane}},\ and\ \bibinfo {author} {\bibfnamefont {B.~A.}\ \bibnamefont {Bernevig}},\ }\bibfield  {title} {\bibinfo {title} {Wallpaper fermions and the nonsymmorphic {Dirac} insulator},\ }\href {https://doi.org/10.1126/science.aan2802} {\bibfield  {journal} {\bibinfo  {journal} {Science}\ }\textbf {\bibinfo {volume} {361}},\ \bibinfo {pages} {246} (\bibinfo {year} {2018})}\BibitemShut {NoStop}%
\bibitem [{\citenamefont {Ale~Crivillero}\ \emph {et~al.}(2023)\citenamefont {Ale~Crivillero}, \citenamefont {Rößler}, \citenamefont {Granovsky}, \citenamefont {Doerr}, \citenamefont {Cook}, \citenamefont {Rosa}, \citenamefont {Müller},\ and\ \citenamefont {Wirth}}]{ale_crivillero_magnetic_2023}%
  \BibitemOpen
  \bibfield  {author} {\bibinfo {author} {\bibfnamefont {M.~V.}\ \bibnamefont {Ale~Crivillero}}, \bibinfo {author} {\bibfnamefont {S.}~\bibnamefont {Rößler}}, \bibinfo {author} {\bibfnamefont {S.}~\bibnamefont {Granovsky}}, \bibinfo {author} {\bibfnamefont {M.}~\bibnamefont {Doerr}}, \bibinfo {author} {\bibfnamefont {M.~S.}\ \bibnamefont {Cook}}, \bibinfo {author} {\bibfnamefont {P.~F.~S.}\ \bibnamefont {Rosa}}, \bibinfo {author} {\bibfnamefont {J.}~\bibnamefont {Müller}},\ and\ \bibinfo {author} {\bibfnamefont {S.}~\bibnamefont {Wirth}},\ }\bibfield  {title} {\bibinfo {title} {Magnetic and electronic properties unveil polaron formation in \ch{Eu5In2Sb6}},\ }\href {https://doi.org/10.1038/s41598-023-28711-z} {\bibfield  {journal} {\bibinfo  {journal} {Scientific Reports}\ }\textbf {\bibinfo {volume} {13}},\ \bibinfo {pages} {1597} (\bibinfo {year} {2023})}\BibitemShut {NoStop}%
\bibitem [{\citenamefont {Crivillero}\ \emph {et~al.}(2022)\citenamefont {Crivillero}, \citenamefont {Rößler}, \citenamefont {Rosa}, \citenamefont {Müller}, \citenamefont {Rößler},\ and\ \citenamefont {Wirth}}]{crivillero_surface_2022}%
  \BibitemOpen
  \bibfield  {author} {\bibinfo {author} {\bibfnamefont {M.~V.~A.}\ \bibnamefont {Crivillero}}, \bibinfo {author} {\bibfnamefont {S.}~\bibnamefont {Rößler}}, \bibinfo {author} {\bibfnamefont {P.~F.~S.}\ \bibnamefont {Rosa}}, \bibinfo {author} {\bibfnamefont {J.}~\bibnamefont {Müller}}, \bibinfo {author} {\bibfnamefont {U.~K.}\ \bibnamefont {Rößler}},\ and\ \bibinfo {author} {\bibfnamefont {S.}~\bibnamefont {Wirth}},\ }\bibfield  {title} {\bibinfo {title} {Surface and electronic structure at atomic length scales of the nonsymmorphic antiferromagnet \ch{Eu5In2Sb6}},\ }\href {https://doi.org/10.1103/PhysRevB.106.035124} {\bibfield  {journal} {\bibinfo  {journal} {Physical Review B}\ }\textbf {\bibinfo {volume} {106}},\ \bibinfo {pages} {035124} (\bibinfo {year} {2022})}\BibitemShut {NoStop}%
\bibitem [{\citenamefont {Tagliati}\ \emph {et~al.}(2012)\citenamefont {Tagliati}, \citenamefont {Krasnov},\ and\ \citenamefont {Rydh}}]{tagliati_differential_2012}%
  \BibitemOpen
  \bibfield  {author} {\bibinfo {author} {\bibfnamefont {S.}~\bibnamefont {Tagliati}}, \bibinfo {author} {\bibfnamefont {V.~M.}\ \bibnamefont {Krasnov}},\ and\ \bibinfo {author} {\bibfnamefont {A.}~\bibnamefont {Rydh}},\ }\bibfield  {title} {\bibinfo {title} {Differential membrane-based nanocalorimeter for high-resolution measurements of low-temperature specific heat},\ }\href {https://doi.org/10.1063/1.4717676} {\bibfield  {journal} {\bibinfo  {journal} {Review of Scientific Instruments}\ }\textbf {\bibinfo {volume} {83}},\ \bibinfo {pages} {055107} (\bibinfo {year} {2012})}\BibitemShut {NoStop}%
\bibitem [{\citenamefont {Wang}\ \emph {et~al.}(2021{\natexlab{b}})\citenamefont {Wang}, \citenamefont {Po}, \citenamefont {Slager},\ and\ \citenamefont {Vishwanath}}]{wang_topological_2021}%
  \BibitemOpen
  \bibfield  {author} {\bibinfo {author} {\bibfnamefont {L.-L.}\ \bibnamefont {Wang}}, \bibinfo {author} {\bibfnamefont {H.~C.}\ \bibnamefont {Po}}, \bibinfo {author} {\bibfnamefont {R.-J.}\ \bibnamefont {Slager}},\ and\ \bibinfo {author} {\bibfnamefont {A.}~\bibnamefont {Vishwanath}},\ }\bibfield  {title} {\bibinfo {title} {Topological descendants of a multicritical {Dirac} semimetal with magnetism and strain},\ }\href {https://doi.org/10.1103/PhysRevB.104.165107} {\bibfield  {journal} {\bibinfo  {journal} {Physical Review B}\ }\textbf {\bibinfo {volume} {104}},\ \bibinfo {pages} {165107} (\bibinfo {year} {2021}{\natexlab{b}})}\BibitemShut {NoStop}%
\bibitem [{\citenamefont {Pohlit}\ \emph {et~al.}(2018)\citenamefont {Pohlit}, \citenamefont {Rößler}, \citenamefont {Ohno}, \citenamefont {Ohno}, \citenamefont {von Molnár}, \citenamefont {Fisk}, \citenamefont {Müller},\ and\ \citenamefont {Wirth}}]{pohlit_evidence_2018}%
  \BibitemOpen
  \bibfield  {author} {\bibinfo {author} {\bibfnamefont {M.}~\bibnamefont {Pohlit}}, \bibinfo {author} {\bibfnamefont {S.}~\bibnamefont {Rößler}}, \bibinfo {author} {\bibfnamefont {Y.}~\bibnamefont {Ohno}}, \bibinfo {author} {\bibfnamefont {H.}~\bibnamefont {Ohno}}, \bibinfo {author} {\bibfnamefont {S.}~\bibnamefont {von Molnár}}, \bibinfo {author} {\bibfnamefont {Z.}~\bibnamefont {Fisk}}, \bibinfo {author} {\bibfnamefont {J.}~\bibnamefont {Müller}},\ and\ \bibinfo {author} {\bibfnamefont {S.}~\bibnamefont {Wirth}},\ }\bibfield  {title} {\bibinfo {title} {Evidence for {Ferromagnetic} {Clusters} in the {Colossal}-{Magnetoresistance} {Material} \ch{EuB6}},\ }\href {https://doi.org/10.1103/PhysRevLett.120.257201} {\bibfield  {journal} {\bibinfo  {journal} {Physical Review Letters}\ }\textbf {\bibinfo {volume} {120}},\ \bibinfo {pages} {257201} (\bibinfo {year} {2018})}\BibitemShut {NoStop}%
\bibitem [{\citenamefont {Das}\ \emph {et~al.}(2012)\citenamefont {Das}, \citenamefont {Amyan}, \citenamefont {Brandenburg}, \citenamefont {Müller}, \citenamefont {Xiong}, \citenamefont {von Molnár},\ and\ \citenamefont {Fisk}}]{das_magnetically_2012}%
  \BibitemOpen
  \bibfield  {author} {\bibinfo {author} {\bibfnamefont {P.}~\bibnamefont {Das}}, \bibinfo {author} {\bibfnamefont {A.}~\bibnamefont {Amyan}}, \bibinfo {author} {\bibfnamefont {J.}~\bibnamefont {Brandenburg}}, \bibinfo {author} {\bibfnamefont {J.}~\bibnamefont {Müller}}, \bibinfo {author} {\bibfnamefont {P.}~\bibnamefont {Xiong}}, \bibinfo {author} {\bibfnamefont {S.}~\bibnamefont {von Molnár}},\ and\ \bibinfo {author} {\bibfnamefont {Z.}~\bibnamefont {Fisk}},\ }\bibfield  {title} {\bibinfo {title} {Magnetically driven electronic phase separation in the semimetallic ferromagnet \ch{EuB6}},\ }\href {https://doi.org/10.1103/PhysRevB.86.184425} {\bibfield  {journal} {\bibinfo  {journal} {Physical Review B}\ }\textbf {\bibinfo {volume} {86}},\ \bibinfo {pages} {184425} (\bibinfo {year} {2012})}\BibitemShut {NoStop}%
\bibitem [{\citenamefont {Süllow}\ \emph {et~al.}(2000)\citenamefont {Süllow}, \citenamefont {Prasad}, \citenamefont {Bogdanovich}, \citenamefont {Aronson}, \citenamefont {Sarrao},\ and\ \citenamefont {Fisk}}]{sullow_magnetotransport_2000}%
  \BibitemOpen
  \bibfield  {author} {\bibinfo {author} {\bibfnamefont {S.}~\bibnamefont {Süllow}}, \bibinfo {author} {\bibfnamefont {I.}~\bibnamefont {Prasad}}, \bibinfo {author} {\bibfnamefont {S.}~\bibnamefont {Bogdanovich}}, \bibinfo {author} {\bibfnamefont {M.~C.}\ \bibnamefont {Aronson}}, \bibinfo {author} {\bibfnamefont {J.~L.}\ \bibnamefont {Sarrao}},\ and\ \bibinfo {author} {\bibfnamefont {Z.}~\bibnamefont {Fisk}},\ }\bibfield  {title} {\bibinfo {title} {Magnetotransport in the low carrier density ferromagnet \ch{EuB6}},\ }\href {https://doi.org/10.1063/1.372460} {\bibfield  {journal} {\bibinfo  {journal} {Journal of Applied Physics}\ }\textbf {\bibinfo {volume} {87}},\ \bibinfo {pages} {5591} (\bibinfo {year} {2000})}\BibitemShut {NoStop}%
\bibitem [{\citenamefont {Zhang}\ \emph {et~al.}(2009)\citenamefont {Zhang}, \citenamefont {Yu}, \citenamefont {von Molnár}, \citenamefont {Fisk},\ and\ \citenamefont {Xiong}}]{zhang_nonlinear_2009}%
  \BibitemOpen
  \bibfield  {author} {\bibinfo {author} {\bibfnamefont {X.}~\bibnamefont {Zhang}}, \bibinfo {author} {\bibfnamefont {L.}~\bibnamefont {Yu}}, \bibinfo {author} {\bibfnamefont {S.}~\bibnamefont {von Molnár}}, \bibinfo {author} {\bibfnamefont {Z.}~\bibnamefont {Fisk}},\ and\ \bibinfo {author} {\bibfnamefont {P.}~\bibnamefont {Xiong}},\ }\bibfield  {title} {\bibinfo {title} {Nonlinear {Hall} {Effect} as a {Signature} of {Electronic} {Phase} {Separation} in the {Semimetallic} {Ferromagnet} \ch{EuB6}},\ }\href {https://doi.org/10.1103/PhysRevLett.103.106602} {\bibfield  {journal} {\bibinfo  {journal} {Physical Review Letters}\ }\textbf {\bibinfo {volume} {103}},\ \bibinfo {pages} {106602} (\bibinfo {year} {2009})}\BibitemShut {NoStop}%
\bibitem [{\citenamefont {Murakawa}\ \emph {et~al.}(2023)\citenamefont {Murakawa}, \citenamefont {Nakaoka}, \citenamefont {Iwase}, \citenamefont {Kida}, \citenamefont {Hagiwara}, \citenamefont {Sakai},\ and\ \citenamefont {Hanasaki}}]{murakawa_giant_2023}%
  \BibitemOpen
  \bibfield  {author} {\bibinfo {author} {\bibfnamefont {H.}~\bibnamefont {Murakawa}}, \bibinfo {author} {\bibfnamefont {Y.}~\bibnamefont {Nakaoka}}, \bibinfo {author} {\bibfnamefont {K.}~\bibnamefont {Iwase}}, \bibinfo {author} {\bibfnamefont {T.}~\bibnamefont {Kida}}, \bibinfo {author} {\bibfnamefont {M.}~\bibnamefont {Hagiwara}}, \bibinfo {author} {\bibfnamefont {H.}~\bibnamefont {Sakai}},\ and\ \bibinfo {author} {\bibfnamefont {N.}~\bibnamefont {Hanasaki}},\ }\bibfield  {title} {\bibinfo {title} {Giant negative magnetoresistance in the layered semiconductor {CeTe}$_{2-x}${Sb}$_x$ with variable magnetic polaron density},\ }\href {https://doi.org/10.1103/PhysRevB.107.165138} {\bibfield  {journal} {\bibinfo  {journal} {Physical Review B}\ }\textbf {\bibinfo {volume} {107}},\ \bibinfo {pages} {165138} (\bibinfo {year} {2023})}\BibitemShut {NoStop}%
\bibitem [{\citenamefont {Manna}\ \emph {et~al.}(2014)\citenamefont {Manna}, \citenamefont {Das}, \citenamefont {de~Souza}, \citenamefont {Schnelle}, \citenamefont {Lang}, \citenamefont {Müller}, \citenamefont {von Molnár},\ and\ \citenamefont {Fisk}}]{manna_lattice_2014}%
  \BibitemOpen
  \bibfield  {author} {\bibinfo {author} {\bibfnamefont {R.~S.}\ \bibnamefont {Manna}}, \bibinfo {author} {\bibfnamefont {P.}~\bibnamefont {Das}}, \bibinfo {author} {\bibfnamefont {M.}~\bibnamefont {de~Souza}}, \bibinfo {author} {\bibfnamefont {F.}~\bibnamefont {Schnelle}}, \bibinfo {author} {\bibfnamefont {M.}~\bibnamefont {Lang}}, \bibinfo {author} {\bibfnamefont {J.}~\bibnamefont {Müller}}, \bibinfo {author} {\bibfnamefont {S.}~\bibnamefont {von Molnár}},\ and\ \bibinfo {author} {\bibfnamefont {Z.}~\bibnamefont {Fisk}},\ }\bibfield  {title} {\bibinfo {title} {Lattice {Strain} {Accompanying} the {Colossal} {Magnetoresistance} {Effect} in \ch{EuB6}},\ }\href {https://doi.org/10.1103/PhysRevLett.113.067202} {\bibfield  {journal} {\bibinfo  {journal} {Physical Review Letters}\ }\textbf {\bibinfo {volume} {113}},\ \bibinfo {pages} {067202} (\bibinfo {year} {2014})}\BibitemShut {NoStop}%
\bibitem [{\citenamefont {Chatterjee}\ \emph {et~al.}(2004)\citenamefont {Chatterjee}, \citenamefont {Yu},\ and\ \citenamefont {Min}}]{chatterjee_spin-polaron_2004}%
  \BibitemOpen
  \bibfield  {author} {\bibinfo {author} {\bibfnamefont {J.}~\bibnamefont {Chatterjee}}, \bibinfo {author} {\bibfnamefont {U.}~\bibnamefont {Yu}},\ and\ \bibinfo {author} {\bibfnamefont {B.~I.}\ \bibnamefont {Min}},\ }\bibfield  {title} {\bibinfo {title} {Spin-polaron model: {Transport} properties of \ch{EuB6}},\ }\href {https://doi.org/10.1103/PhysRevB.69.134423} {\bibfield  {journal} {\bibinfo  {journal} {Physical Review B}\ }\textbf {\bibinfo {volume} {69}},\ \bibinfo {pages} {134423} (\bibinfo {year} {2004})}\BibitemShut {NoStop}%
\bibitem [{\citenamefont {Slater}(1951)}]{slater_magnetic_1951}%
  \BibitemOpen
  \bibfield  {author} {\bibinfo {author} {\bibfnamefont {J.~C.}\ \bibnamefont {Slater}},\ }\bibfield  {title} {\bibinfo {title} {Magnetic {Effects} and the {Hartree}-{Fock} {Equation}},\ }\href {https://doi.org/10.1103/PhysRev.82.538} {\bibfield  {journal} {\bibinfo  {journal} {Physical Review}\ }\textbf {\bibinfo {volume} {82}},\ \bibinfo {pages} {538} (\bibinfo {year} {1951})}\BibitemShut {NoStop}%
\bibitem [{\citenamefont {Tomioka}\ \emph {et~al.}(1997)\citenamefont {Tomioka}, \citenamefont {Asamitsu}, \citenamefont {Kuwahara}, \citenamefont {Moritomo}, \citenamefont {Kasai}, \citenamefont {Kumai},\ and\ \citenamefont {Tokura}}]{tomioka_magnetic-field-induced_1997}%
  \BibitemOpen
  \bibfield  {author} {\bibinfo {author} {\bibfnamefont {Y.}~\bibnamefont {Tomioka}}, \bibinfo {author} {\bibfnamefont {A.}~\bibnamefont {Asamitsu}}, \bibinfo {author} {\bibfnamefont {H.}~\bibnamefont {Kuwahara}}, \bibinfo {author} {\bibfnamefont {Y.}~\bibnamefont {Moritomo}}, \bibinfo {author} {\bibfnamefont {M.}~\bibnamefont {Kasai}}, \bibinfo {author} {\bibfnamefont {R.}~\bibnamefont {Kumai}},\ and\ \bibinfo {author} {\bibfnamefont {Y.}~\bibnamefont {Tokura}},\ }\bibfield  {title} {\bibinfo {title} {Magnetic-field-induced metal-insulator transition in perovskite-type manganese oxides},\ }\href {https://doi.org/10.1016/S0921-4526(97)00013-6} {\bibfield  {journal} {\bibinfo  {journal} {Physica B: Condensed Matter}\ }\bibinfo {series} {Proceedings of the {Yamada} {Conference} {XLV}, the {International} {Conference} on the {Physics} of {Transition} {Metals}},\ \textbf {\bibinfo {volume} {237-238}},\ \bibinfo {pages} {6} (\bibinfo {year} {1997})}\BibitemShut {NoStop}%
\bibitem [{\citenamefont {Tomioka}\ \emph {et~al.}(1995)\citenamefont {Tomioka}, \citenamefont {Asamitsu}, \citenamefont {Moritomo}, \citenamefont {Kuwahara},\ and\ \citenamefont {Tokura}}]{tomioka_collapse_1995}%
  \BibitemOpen
  \bibfield  {author} {\bibinfo {author} {\bibfnamefont {Y.}~\bibnamefont {Tomioka}}, \bibinfo {author} {\bibfnamefont {A.}~\bibnamefont {Asamitsu}}, \bibinfo {author} {\bibfnamefont {Y.}~\bibnamefont {Moritomo}}, \bibinfo {author} {\bibfnamefont {H.}~\bibnamefont {Kuwahara}},\ and\ \bibinfo {author} {\bibfnamefont {Y.}~\bibnamefont {Tokura}},\ }\bibfield  {title} {\bibinfo {title} {Collapse of a {Charge}-{Ordered} {State} under a {Magnetic} {Field} in {Pr}$_\frac{1}{2}${Sr}$_\frac{1}{2}${MnO}$_3$},\ }\href {https://doi.org/10.1103/PhysRevLett.74.5108} {\bibfield  {journal} {\bibinfo  {journal} {Physical Review Letters}\ }\textbf {\bibinfo {volume} {74}},\ \bibinfo {pages} {5108} (\bibinfo {year} {1995})}\BibitemShut {NoStop}%
\bibitem [{\citenamefont {Rao}(2000)}]{rao_charge_2000}%
  \BibitemOpen
  \bibfield  {author} {\bibinfo {author} {\bibfnamefont {C.~N.~R.}\ \bibnamefont {Rao}},\ }\bibfield  {title} {\bibinfo {title} {Charge, {Spin}, and {Orbital} {Ordering} in the {Perovskite} {Manganates}, {Ln}$_{1-x}${A}$_x${MnO}$_3$ ({Ln} = {Rare} {Earth}, {A} = {Ca} or {Sr})},\ }\href {https://doi.org/10.1021/jp0004866} {\bibfield  {journal} {\bibinfo  {journal} {The Journal of Physical Chemistry B}\ }\textbf {\bibinfo {volume} {104}},\ \bibinfo {pages} {5877} (\bibinfo {year} {2000})}\BibitemShut {NoStop}%
\bibitem [{\citenamefont {Li}\ \emph {et~al.}()\citenamefont {Li}, \citenamefont {Been}, \citenamefont {Balguri}, \citenamefont {Jia}, \citenamefont {Mahendru}, \citenamefont {Wang}, \citenamefont {Cui}, \citenamefont {Chen}, \citenamefont {Hashimoto}, \citenamefont {Lu}, \citenamefont {Moritz}, \citenamefont {Zaanen}, \citenamefont {Tafti}, \citenamefont {Devereaux},\ and\ \citenamefont {Shen}}]{li_colossal}%
  \BibitemOpen
  \bibfield  {author} {\bibinfo {author} {\bibfnamefont {Y.-F.}\ \bibnamefont {Li}}, \bibinfo {author} {\bibfnamefont {E.~M.}\ \bibnamefont {Been}}, \bibinfo {author} {\bibfnamefont {S.}~\bibnamefont {Balguri}}, \bibinfo {author} {\bibfnamefont {C.-J.}\ \bibnamefont {Jia}}, \bibinfo {author} {\bibfnamefont {M.~B.}\ \bibnamefont {Mahendru}}, \bibinfo {author} {\bibfnamefont {Z.-C.}\ \bibnamefont {Wang}}, \bibinfo {author} {\bibfnamefont {Y.}~\bibnamefont {Cui}}, \bibinfo {author} {\bibfnamefont {S.-D.}\ \bibnamefont {Chen}}, \bibinfo {author} {\bibfnamefont {M.}~\bibnamefont {Hashimoto}}, \bibinfo {author} {\bibfnamefont {D.-H.}\ \bibnamefont {Lu}}, \bibinfo {author} {\bibfnamefont {B.}~\bibnamefont {Moritz}}, \bibinfo {author} {\bibfnamefont {J.}~\bibnamefont {Zaanen}}, \bibinfo {author} {\bibfnamefont {F.}~\bibnamefont {Tafti}}, \bibinfo {author} {\bibfnamefont {T.~P.}\ \bibnamefont {Devereaux}},\ and\ \bibinfo {author} {\bibfnamefont {Z.-X.}\ \bibnamefont {Shen}},\ }\bibfield  {title} {\bibinfo {title}
  {Colossal magnetoresistance from spin-polarized polarons in an ising system},\ }\href {https://doi.org/10.1073/pnas.2409846121} {\bibfield  {journal} {\bibinfo  {journal} {Proceedings of the National Academy of Sciences}\ }\textbf {\bibinfo {volume} {121}},\ \bibinfo {pages} {e2409846121}}\BibitemShut {NoStop}%
\bibitem [{\citenamefont {Lai}\ \emph {et~al.}(2010)\citenamefont {Lai}, \citenamefont {Nakamura}, \citenamefont {Kundhikanjana}, \citenamefont {Kawasaki}, \citenamefont {Tokura}, \citenamefont {Kelly},\ and\ \citenamefont {Shen}}]{lai_mesoscopic_2010}%
  \BibitemOpen
  \bibfield  {author} {\bibinfo {author} {\bibfnamefont {K.}~\bibnamefont {Lai}}, \bibinfo {author} {\bibfnamefont {M.}~\bibnamefont {Nakamura}}, \bibinfo {author} {\bibfnamefont {W.}~\bibnamefont {Kundhikanjana}}, \bibinfo {author} {\bibfnamefont {M.}~\bibnamefont {Kawasaki}}, \bibinfo {author} {\bibfnamefont {Y.}~\bibnamefont {Tokura}}, \bibinfo {author} {\bibfnamefont {M.~A.}\ \bibnamefont {Kelly}},\ and\ \bibinfo {author} {\bibfnamefont {Z.-X.}\ \bibnamefont {Shen}},\ }\bibfield  {title} {\bibinfo {title} {Mesoscopic {Percolating} {Resistance} {Network} in a {Strained} {Manganite} {Thin} {Film}},\ }\href {https://doi.org/10.1126/science.1189925} {\bibfield  {journal} {\bibinfo  {journal} {Science}\ }\textbf {\bibinfo {volume} {329}},\ \bibinfo {pages} {190} (\bibinfo {year} {2010})}\BibitemShut {NoStop}%
\bibitem [{\citenamefont {Rahn}\ \emph {et~al.}(2018)\citenamefont {Rahn}, \citenamefont {Soh}, \citenamefont {Francoual}, \citenamefont {Veiga}, \citenamefont {Strempfer}, \citenamefont {Mardegan}, \citenamefont {Yan}, \citenamefont {Guo}, \citenamefont {Shi},\ and\ \citenamefont {Boothroyd}}]{rahn_coupling_2018}%
  \BibitemOpen
  \bibfield  {author} {\bibinfo {author} {\bibfnamefont {M.~C.}\ \bibnamefont {Rahn}}, \bibinfo {author} {\bibfnamefont {J.-R.}\ \bibnamefont {Soh}}, \bibinfo {author} {\bibfnamefont {S.}~\bibnamefont {Francoual}}, \bibinfo {author} {\bibfnamefont {L.~S.~I.}\ \bibnamefont {Veiga}}, \bibinfo {author} {\bibfnamefont {J.}~\bibnamefont {Strempfer}}, \bibinfo {author} {\bibfnamefont {J.}~\bibnamefont {Mardegan}}, \bibinfo {author} {\bibfnamefont {D.~Y.}\ \bibnamefont {Yan}}, \bibinfo {author} {\bibfnamefont {Y.~F.}\ \bibnamefont {Guo}}, \bibinfo {author} {\bibfnamefont {Y.~G.}\ \bibnamefont {Shi}},\ and\ \bibinfo {author} {\bibfnamefont {A.~T.}\ \bibnamefont {Boothroyd}},\ }\bibfield  {title} {\bibinfo {title} {Coupling of magnetic order and charge transport in the candidate {Dirac} semimetal \ch{EuCd2As2}},\ }\href {https://doi.org/10.1103/PhysRevB.97.214422} {\bibfield  {journal} {\bibinfo  {journal} {Physical Review B}\ }\textbf {\bibinfo {volume} {97}},\ \bibinfo {pages} {214422} (\bibinfo {year}
  {2018})}\BibitemShut {NoStop}%
\bibitem [{\citenamefont {Nasrallah}\ \emph {et~al.}()\citenamefont {Nasrallah}, \citenamefont {Santos-Cottin}, \citenamefont {Le~Mardelé}, \citenamefont {Mohelský}, \citenamefont {Wyzula}, \citenamefont {Akšamović}, \citenamefont {Sačer}, \citenamefont {Barrett}, \citenamefont {Galloway}, \citenamefont {Rigaux}, \citenamefont {Guo}, \citenamefont {Puppin}, \citenamefont {Živković}, \citenamefont {Dil}, \citenamefont {Novak}, \citenamefont {Homes}, \citenamefont {Orlita}, \citenamefont {Barišić},\ and\ \citenamefont {Akrap}}]{nasrallah_magneto-optical_2024}%
  \BibitemOpen
  \bibfield  {author} {\bibinfo {author} {\bibfnamefont {S.}~\bibnamefont {Nasrallah}}, \bibinfo {author} {\bibfnamefont {D.}~\bibnamefont {Santos-Cottin}}, \bibinfo {author} {\bibfnamefont {F.}~\bibnamefont {Le~Mardelé}}, \bibinfo {author} {\bibfnamefont {I.}~\bibnamefont {Mohelský}}, \bibinfo {author} {\bibfnamefont {J.}~\bibnamefont {Wyzula}}, \bibinfo {author} {\bibfnamefont {L.}~\bibnamefont {Akšamović}}, \bibinfo {author} {\bibfnamefont {P.}~\bibnamefont {Sačer}}, \bibinfo {author} {\bibfnamefont {J.~W.~H.}\ \bibnamefont {Barrett}}, \bibinfo {author} {\bibfnamefont {W.}~\bibnamefont {Galloway}}, \bibinfo {author} {\bibfnamefont {K.}~\bibnamefont {Rigaux}}, \bibinfo {author} {\bibfnamefont {F.}~\bibnamefont {Guo}}, \bibinfo {author} {\bibfnamefont {M.}~\bibnamefont {Puppin}}, \bibinfo {author} {\bibfnamefont {I.}~\bibnamefont {Živković}}, \bibinfo {author} {\bibfnamefont {J.~H.}\ \bibnamefont {Dil}}, \bibinfo {author} {\bibfnamefont {M.}~\bibnamefont {Novak}}, \bibinfo {author} {\bibfnamefont
  {C.~C.}\ \bibnamefont {Homes}}, \bibinfo {author} {\bibfnamefont {M.}~\bibnamefont {Orlita}}, \bibinfo {author} {\bibfnamefont {N.}~\bibnamefont {Barišić}},\ and\ \bibinfo {author} {\bibfnamefont {A.}~\bibnamefont {Akrap}},\ }\bibfield  {title} {\bibinfo {title} {Magneto-optical response of the magnetic semiconductors \ch{EuCd2X2\ ( X =P, As, Sb)}},\ }\href {https://doi.org/10.1103/PhysRevB.110.L201201} {\bibfield  {journal} {\bibinfo  {journal} {Physical Review B}\ }\textbf {\bibinfo {volume} {110}},\ \bibinfo {pages} {L201201}}\BibitemShut {NoStop}%
\bibitem [{\citenamefont {Chen}\ \emph {et~al.}(2024)\citenamefont {Chen}, \citenamefont {Wang}, \citenamefont {Zhou}, \citenamefont {Yang}, \citenamefont {Liu}, \citenamefont {Lu}, \citenamefont {Ren}, \citenamefont {Cao}, \citenamefont {Tafti}, \citenamefont {Dong},\ and\ \citenamefont {Wang}}]{chen_manipulating_2024}%
  \BibitemOpen
  \bibfield  {author} {\bibinfo {author} {\bibfnamefont {X.}~\bibnamefont {Chen}}, \bibinfo {author} {\bibfnamefont {Z.}~\bibnamefont {Wang}}, \bibinfo {author} {\bibfnamefont {Z.}~\bibnamefont {Zhou}}, \bibinfo {author} {\bibfnamefont {W.}~\bibnamefont {Yang}}, \bibinfo {author} {\bibfnamefont {Y.}~\bibnamefont {Liu}}, \bibinfo {author} {\bibfnamefont {J.-Y.}\ \bibnamefont {Lu}}, \bibinfo {author} {\bibfnamefont {Z.}~\bibnamefont {Ren}}, \bibinfo {author} {\bibfnamefont {G.-H.}\ \bibnamefont {Cao}}, \bibinfo {author} {\bibfnamefont {F.}~\bibnamefont {Tafti}}, \bibinfo {author} {\bibfnamefont {S.}~\bibnamefont {Dong}},\ and\ \bibinfo {author} {\bibfnamefont {Z.-C.}\ \bibnamefont {Wang}},\ }\bibfield  {title} {\bibinfo {title} {Manipulating magnetism and transport properties of \ch{EuCd2P2} with a low carrier concentration},\ }\href {https://doi.org/10.1103/PhysRevB.109.224428} {\bibfield  {journal} {\bibinfo  {journal} {Physical Review B}\ }\textbf {\bibinfo {volume} {109}},\ \bibinfo {pages} {224428}
  (\bibinfo {year} {2024})}\BibitemShut {NoStop}%
\end{thebibliography}%

\end{document}



\title{Supplemental Material:\\Two types of colossal magnetoresistance with distinct mechanisms in \ch{Eu5In2As6}} 

\author{Sudhaman~Balguri$^\dagger$}
\affiliation{Department of Physics, Boston College, Chestnut Hill, MA 02467, USA}
\thanks{These authors contributed equally to this work.}

\author{Mira~B.~Mahendru$^\dagger$}
\affiliation{Department of Physics, Boston College, Chestnut Hill, MA 02467, USA}

\author{Enrique~O.~González~Delgado}
\affiliation{Department of Physics, Boston College, Chestnut Hill, MA 02467, USA}

\author{Kyle~Fruhling}
\affiliation{Department of Physics, Boston College, Chestnut Hill, MA 02467, USA}

\author{Xiaohan~Yao}
\affiliation{Department of Physics, Boston College, Chestnut Hill, MA 02467, USA}

\author{David~E.~Graf}
\affiliation{National High Magnetic Field Laboratory, Tallahassee, Florida 32310, USA}

\author{Jose~A.~Rodriguez-Rivera}
\affiliation{NIST Center for Neutron Research, National Institute of Standards and Technology, Gaithersburg, Maryland 20899, USA}
\affiliation{Department of Materials Science and Eng., University of Maryland, College Park, MD 20742-2115}

\author{Adam~A.~Aczel}
\affiliation{Neutron Scattering Division, Oak Ridge National Laboratory, Oak Ridge, Tennessee 37831, USA}

\author{Andreas~Rydh}
\affiliation{Department of Physics, Stockholm University, 106 91 Stockholm, Sweden}

\author{Jonathan~Gaudet}
\affiliation{NIST Center for Neutron Research, National Institute of Standards and Technology, Gaithersburg, Maryland 20899, USA}
\affiliation{Department of Materials Science and Eng., University of Maryland, College Park, MD 20742-2115}

\author{Fazel~Tafti}
\email{fazel.tafti@bc.edu}
\affiliation{Department of Physics, Boston College, Chestnut Hill, MA 02467, USA}


\maketitle


\section{\label{sec:structure}Crystal Growth and X-ray Diffraction}
\begin{figure}
  \includegraphics[width=0.46\textwidth]{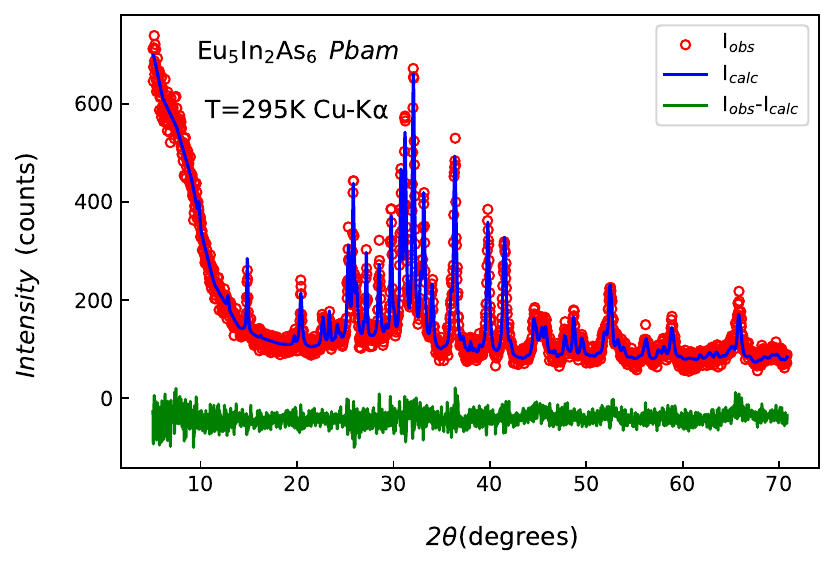}
  \caption{\label{fig:PXRD}
  Rietveld refinement of the PXRD pattern of \ch{Eu5In2As6} in space group $Pbam$ (No. 55).
  The red, blue, and green curves correspond to the observed, calculated, and difference intensities, respectively, while the black ticks mark the expected positions of the Bragg peaks.  
  }
\end{figure}
Large crystals of \ch{Eu5In2As6} were grown using the flux method.
The crystals are shiny and pillar-shaped with the longest dimension along the c-axis as shown in the main text Fig.~1.
They are stable in air but undergo hydrolysis when exposed to moisture, releasing arsine gas and leaving a red powder.
It is advisable to use a fume hood while polishing the samples to avoid potential arsine poisoning.

Powder x-ray diffraction (PXRD) was used to determine the crystal structure of \ch{Eu5In2As6}.
The Rietveld refinement is shown in Fig.~\ref{fig:PXRD}, where the observed Bragg peaks appear in red and the calculated pattern in blue.
The calculated intensities correspond to an orthorhombic unit cell in the space group $Pbam$ with three Eu, one In, and three As sites as summarized in Table~\ref{tab:wyckoff}.
The structural parameters and fit quality factors are summarized in Table~\ref{tab:rietveld}.

\begin{table}
  \caption{\label{tab:wyckoff}Atomic coordinates, site occupancies, and isotropic Debye-Waller factors from PXRD Rietveld refinement of \ch{Eu5In2As6} in space group $Pbam$ at room temperature.
  }
   \begin{tabular}{|ccccccc|}
   \hline
   atom   & site &  $x$    &    $y$    &   $z$      &   occ.       &    $B_{\textrm{iso}}$~(\AA$^2$)   \\
   \hline    
  Eu1     & 4g   & 0.3250  &  0.0000  &  0.00000  &  0.500  &  1.0   \\ 
  Eu2     & 4g   & 0.0853  &  0.2529  &  0.00000  &  0.500  &  1.0   \\
  Eu3     & 2a   & 0.0000  &  0.0000  &  0.00000  &  0.250  &  1.0   \\ 
  In1     & 4h   & 0.3291  &  0.2160  &  0.50000  &  0.500  &  1.0   \\ 
  As1     & 4g   & 0.3364  &  0.3247  &  0.00000  &  0.500  &  1.5   \\ 
  As2     & 4h   & 0.1532  &  0.0987  &  0.50000  &  0.500  &  1.5   \\ 
  As3     & 4h   & 0.0208  &  0.4105  &  0.50000  &  0.500  &  1.5   \\ 
   \hline
  \end{tabular}
\end{table}
\begin{table}
  \caption{\label{tab:rietveld}Unit cell parameters of \ch{Eu5In2As6} and quality factors of the PXRD Rietveld refinement at room temperature.
  }
  \begin{tabular}{|cc|cc|}
  \hline
  \multicolumn{2}{|c|}{ Unit cell parameters }  &       \multicolumn{2}{c|}{ Refinement parameters } \\
  \hline
   Space Group              & $Pbam$            &  Parameters                     & 20               \\
   $a$  (\AA)               & $11.8839(5)$    & $R_{\mathrm {Bragg}}$ (\%)      & 15.0             \\
   $b$  (\AA)               & $13.7885(6)$    & $R_{\mathrm {F}}$ (\%)          & 18.4             \\
   $c$ (\AA)                & $4.3506(2)$     & $R_{\mathrm {exp}}$ (\%)        & 29.1             \\
   V (\AA$^3$)              & $712.909$         & $R_{\mathrm {p}}$ (\%)          & 37.8             \\
   $Z$                      & $2$               & $R_{\mathrm {wp}}$ (\%)         & 33.1             \\
   $\rho$~(gr\,cm$^{-3}$)   & $6.703$           & $\chi^2$                        & 1.30             \\
   \hline
   \end{tabular}
\end{table}

\section{\label{sec:curie}Curie-Weiss analysis}
\begin{figure}
  \includegraphics[width=0.46\textwidth]{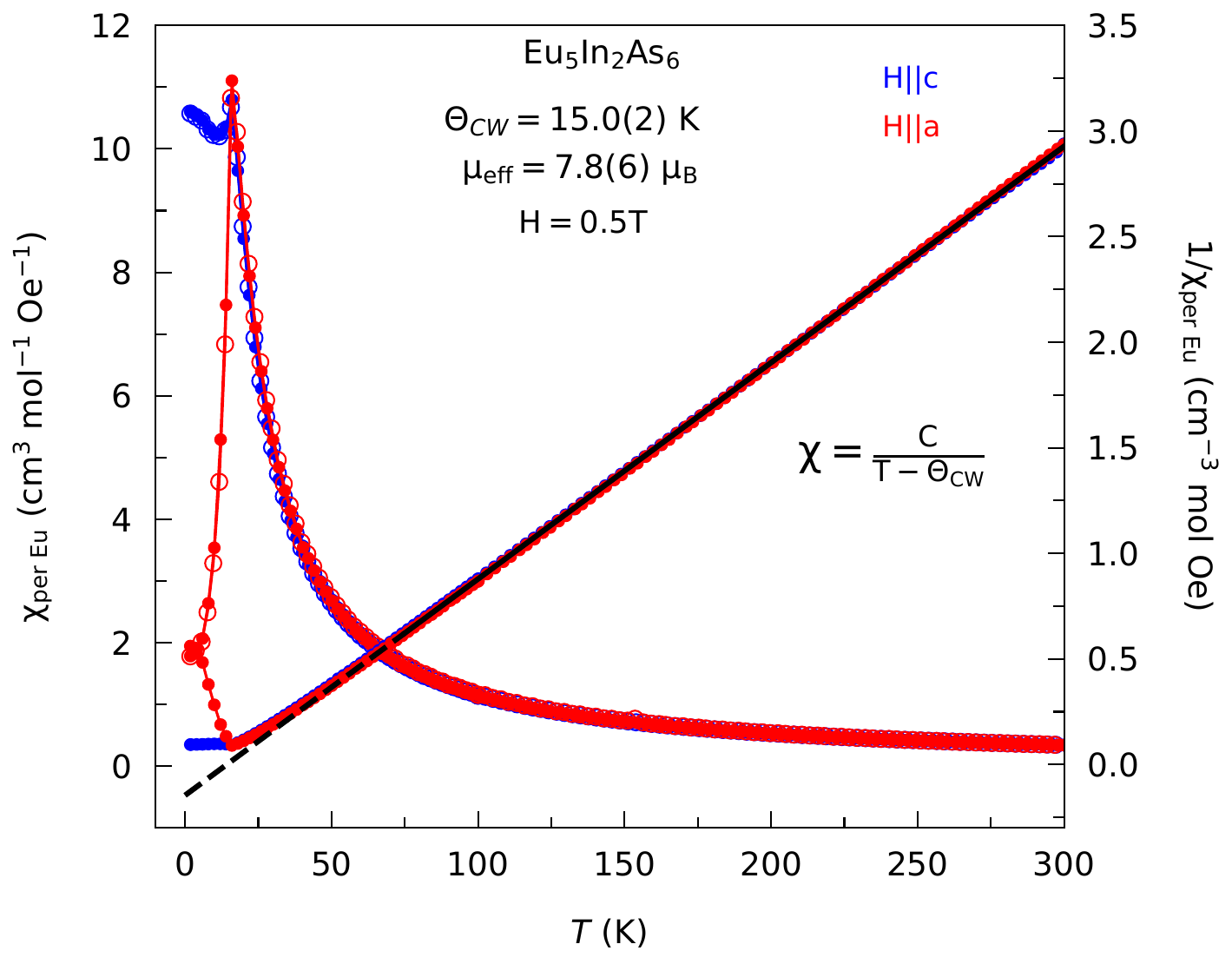}
  \caption{\label{fig:CW}
 Curie-Weiss analysis on the AC susceptibility data for both in-plane and out-of-plane field directions.
  }
\end{figure}
The Curie-Weiss (CW) analysis is shown in Fig.~\ref{fig:CW} for both $H\|a$ (red) and $H\|c$ (blue) directions.
Note that the high-temperature tail of the magnetic susceptibility is nearly identical for both field directions, so the same fit goes through both data sets.
It yields the Curie-Weiss temperature $\Theta_\text{CW}=15.0(2)$~K and effective moment $\mu_\text{eff}=7.8(6)$~\ub.
Despite the AFM order at $T_\text{N1}=16.0(2)$~K, the CW temperature is positive due to the in-plane FM correlations among Eu$^{2+}$ moments (see Fig.~1e,f in the main text).
The absence of a splitting between the field-cooled (FC) and zero-field-cooled (ZFC) data, shown respectively by full and empty symbols in Fig.~\ref{fig:CW}, confirms AFM ordering.

\section{\label{sec:specific heat}Heat capacity}
\begin{figure}
  \includegraphics[width=0.46\textwidth]{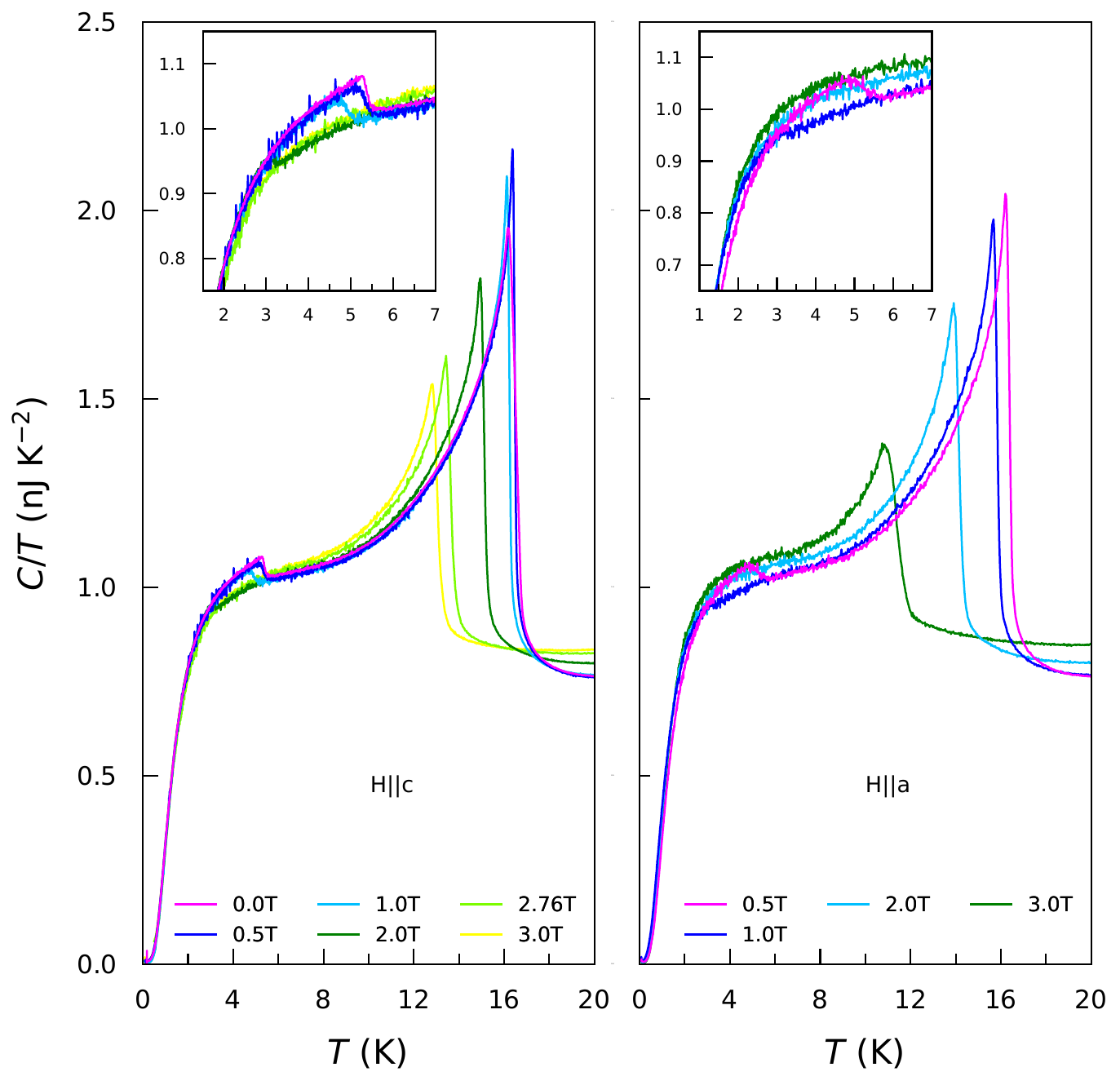}
  \caption{\label{fig:HC_Andreas}
 Heat capacity data obtain by a nanocalorimeter from 20 to 0.05~K.
 Both $T_\text{N1}$ and $T_\text{N2}$ transitions are suppressed by an external magnetic field.
 These data were particularly helpful in mapping the $T_\text{N2}$ phase boundary (Figs. 2c,f in the main text).
  }
\end{figure}
In the main text, the anharmonic Debye model was used to subtract the lattice contribution to the heat capacity.
Before using such a model, several attempts were made to grow crystals of \ch{Sr5In2As6} and \ch{Ba5In2As6}, which could have been used as a lattice model for the title compound.
These attempts were unsuccessful, hence a model calculation was used for the phonon background subtraction.
The simplest expression for the heat capacity in the anharmonic Debye model involving only nearest neighbor interactions in a body-centered cubic lattice is
\begin{equation}
\label{eq:anharmonic}
C=9R\left(\frac{T}{\Theta_\text{D}}\right)^3 \int_0^{\frac{\Theta_\text{D}}{T}} \frac{x^4e^x dx}{(e^x-1)^2~(1+Ax)}
\end{equation}
where $A$ is a constant and $x=\hbar ck/k_BT$ (Stern \emph{et al.} [29]).
The black dashed line in the main Fig.~1b represents the heat capacity calculated by equation ~\ref{eq:anharmonic}. 

All the ``raw'' specific heat data used for mapping the phase diagram of \ch{Eu5In2As6} were shown in the main text Fig.~2, except for a subset of data that was necessary to map the AFM-0 phase.
It was necessary to use a nanocalorimeter at millikelvin temperatures to obtain the data in Fig.~\ref{fig:HC_Andreas}

\section{\label{sec:TDO}Tunnel Diode Oscillator (TDO)}
\begin{figure}
  \includegraphics[width=0.46\textwidth]{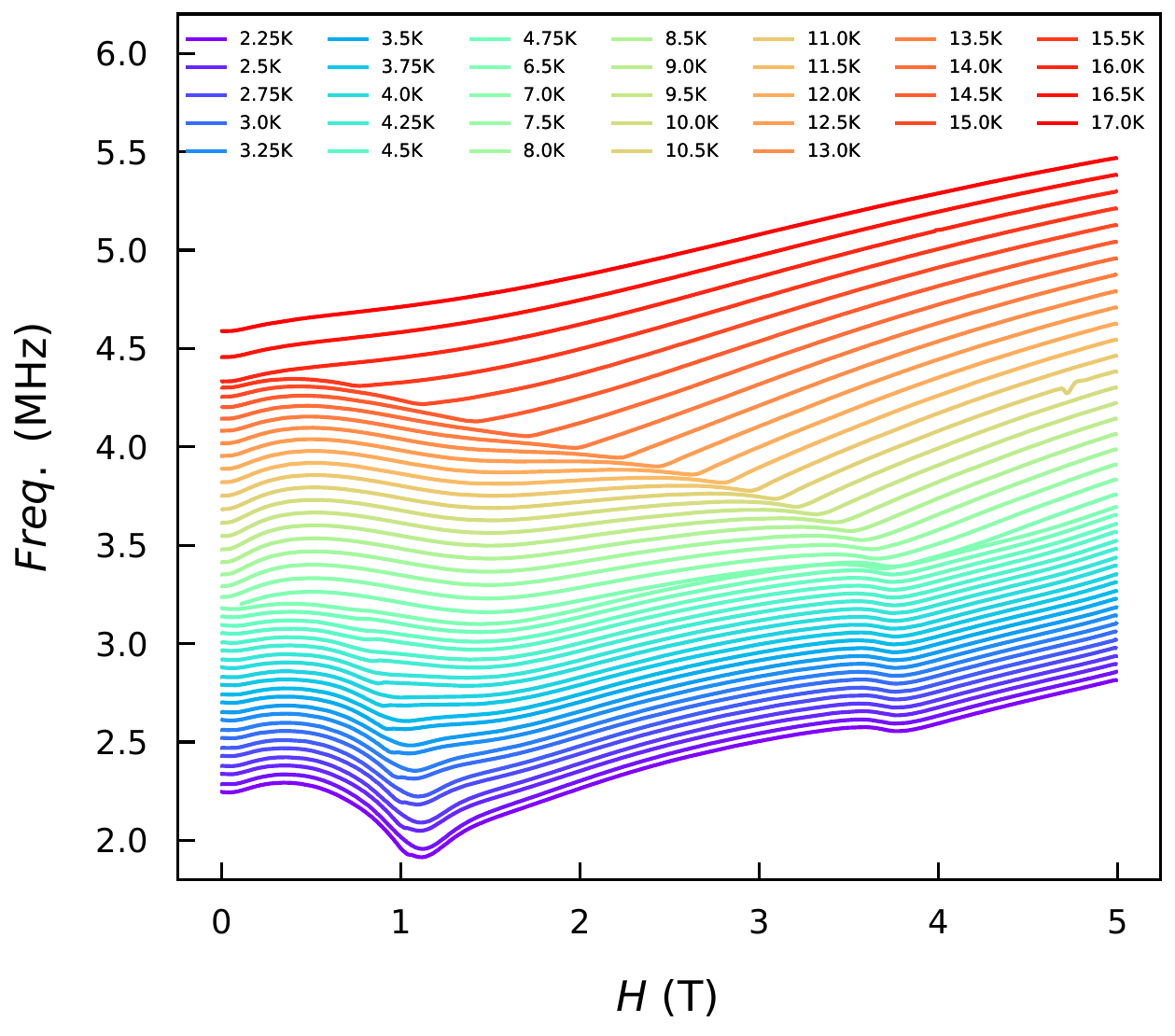}
  \caption{\label{fig:TDO_FH_Eu526_offset}
 Field dependency in TDO frequency at different temperatures.
  }
\end{figure}
\begin{figure}
  \includegraphics[width=0.46\textwidth]{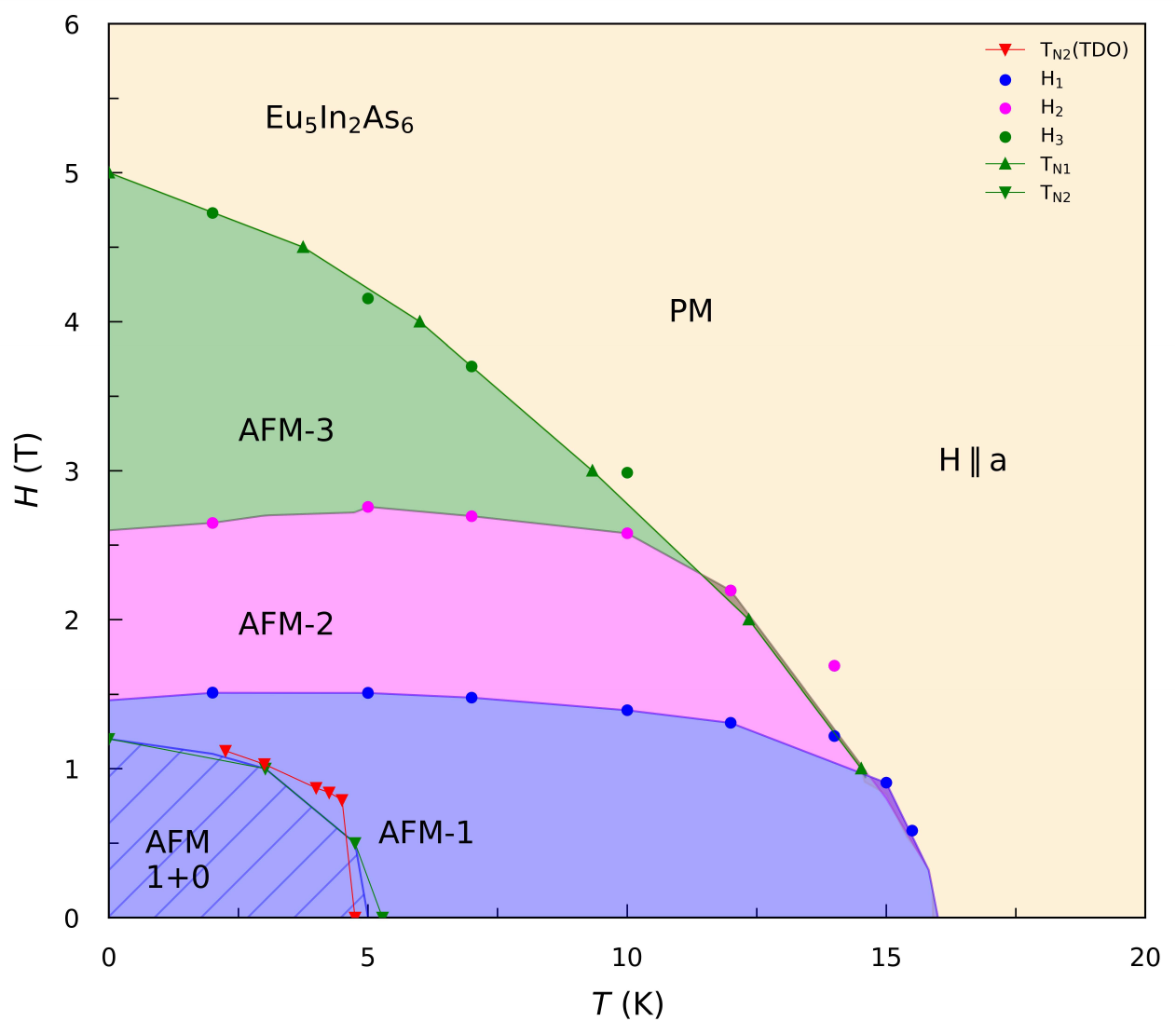}
  \caption{\label{fig:TDO_phasediagram}
 Phase diagram including the TDO data.
  }
\end{figure}

We used TDO to confirm the boundaries of the phase-separated region, AFM-1 + 0, in the phase diagram of \ch{Eu5In2As6}.
Figure~\ref{fig:TDO_FH_Eu526_offset} shows the changes in the TDO frequency at different temperatures.
By observing the changes in the TDO signal we can map, to a good extent, the $H_3$ and $T_{N2}$ components of the phase diagram as shown in Fig.~\ref{fig:TDO_phasediagram}.
The red down-triangles from the TDO measurements confirm the boundaries of the AFM-1+0 phase determined from heat capacity measurements (green down-triangles).

Tunnel diode oscillator (TDO) measurements were performed with the sample coil being a part of an oscillator circuit at a resonant frequency of 65~MHz. 
The TDO circuit board and inductor containing the sample were cooled to low temperatures while a heterodyne circuit at room temperature amplified and mixed the RF signal for recording with a frequency counter.

\section{\label{sec:SAMPLES}Sample dependence of the resistivity and Hall effect}
\begin{figure}
  \includegraphics[width=0.46\textwidth]{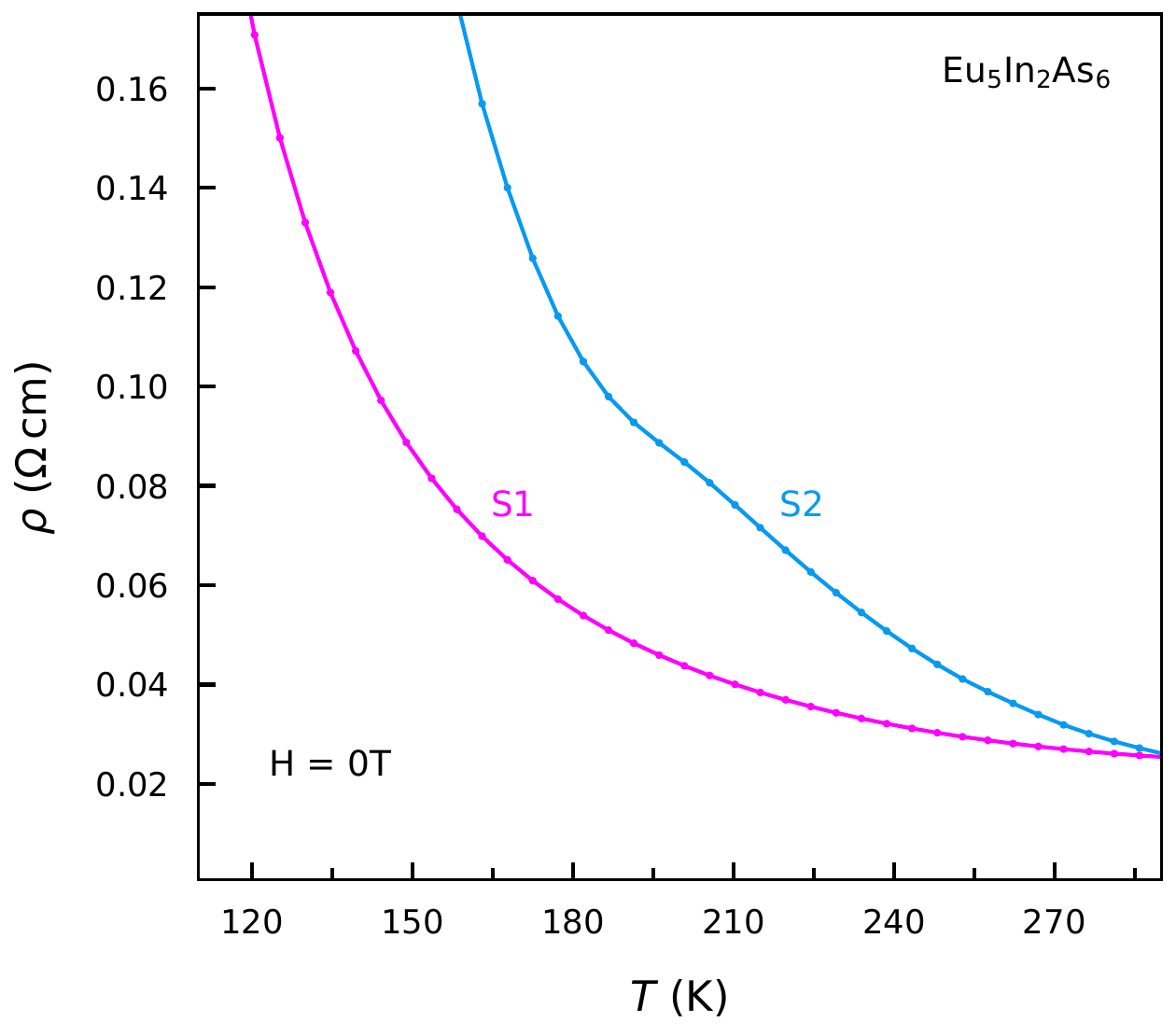}
  \caption{\label{fig:S123_RT}
 Sample dependence of the high-temperature resistivity.
  }
\end{figure}
\begin{figure}
  \includegraphics[width=0.46\textwidth]{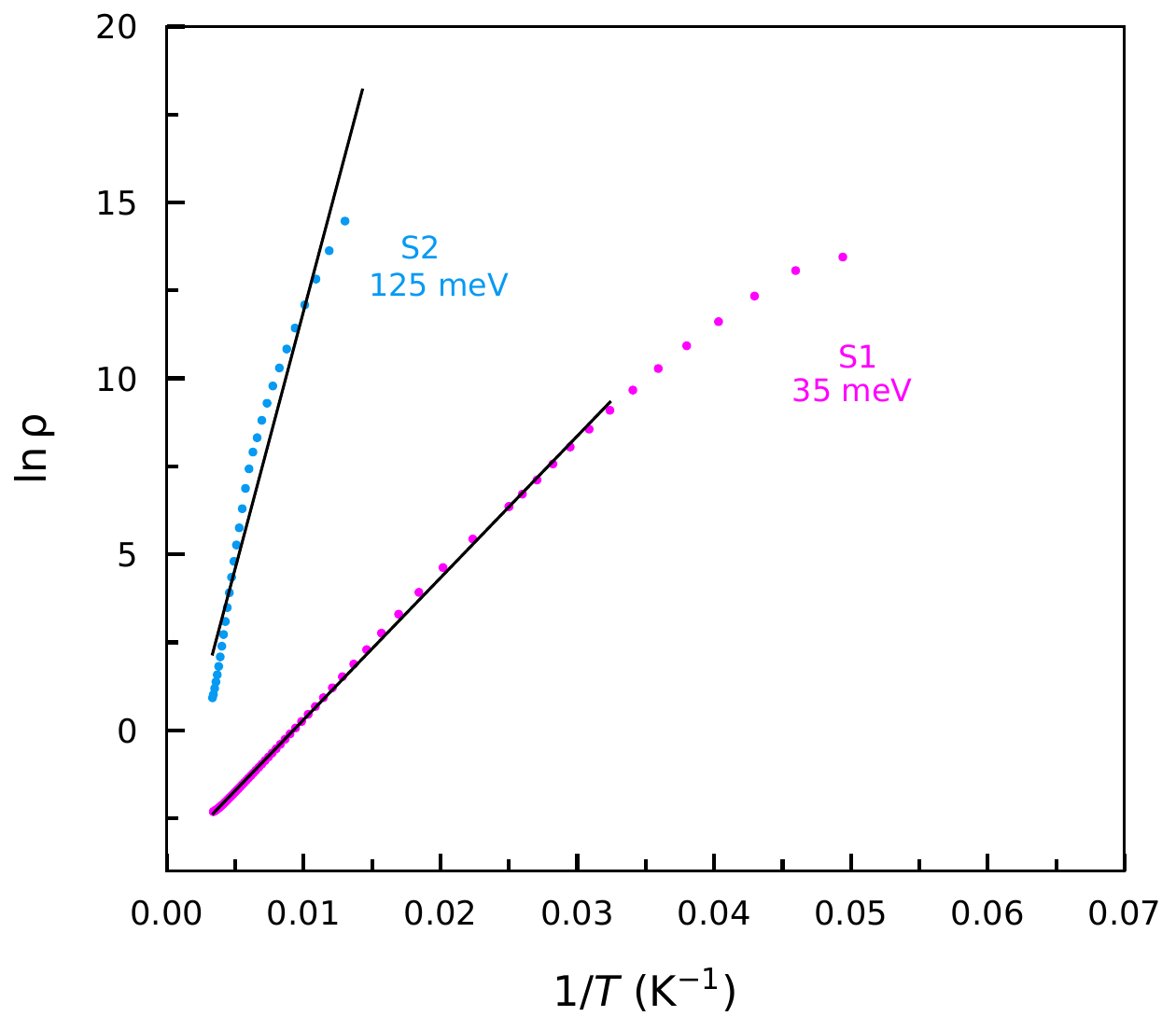}
  \caption{\label{fig:S123_GAP}
 Arrhenius analysis on the resistivity data from samples S1 and S2.
  }
\end{figure}
\begin{figure}
  \includegraphics[width=0.46\textwidth]{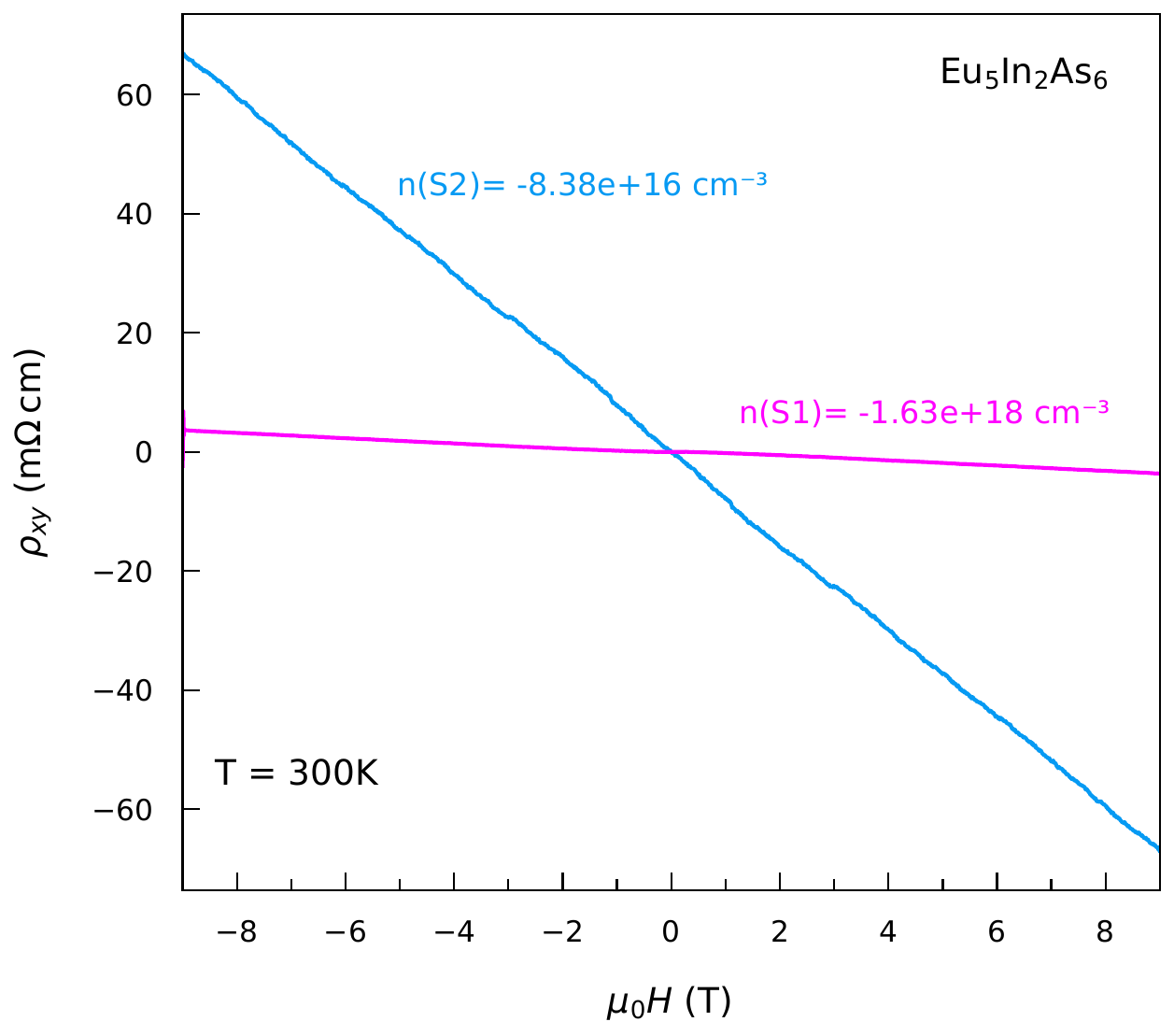}
  \caption{\label{fig:S123_Hall}
 Hall resistance as function of field in samples S1 and S3 at room temperature.
  }
\end{figure}
Figure~\ref{fig:S123_RT} shows that the high-temperature resistivity $\rho(T)$ of different \ch{Eu5In2As6} samples could have different activation gaps.
We extracted the gap values in samples S1 and S2 using an Arrhenius analysis as shown in Fig.~\ref{fig:S123_GAP}.
We also measured Hall effect on both samples (Fig.~\ref{fig:S123_Hall}) and found a smaller carrier density in sample S2 that has a larger gap.
The carrier concentration was extracted using the single band relationship $R_H=1/ne$ where $R_H$, $n$, and $e$ are the Hall coefficient, carrier concentration, and electron charge.
From here, we estimate a concentration of $1.6\times10^{18}$~cm$^{-3}$ and $8.4\times10^{16}$~cm$^{-3}$ for n-type carriers in samples S1 and S2, respectively.

Such variations in the activation gap and carrier concentration in different samples suggest that \ch{Eu5In2As6} is a self-doped semiconductor with extrinsic carriers.
These carriers are coming from the slight off-stoichiometry of Eu and As, confirmed by energy dispersive x-ray spectroscopy (EDX).
The EDX measurements on 6 spots in 2 samples gave the average composition of Eu$_{4.8}$In$_{1.9}$As$_{6.3}$ with approximately 10\% error for each element. 
The measurements were performed in an FEI Scios DualBeam electron microscope equipped with Oxford detector.

\section{\label{sec:SCALING}Spin Fluctuation Scaling Analysis}
In this work, we argue that mangetic polarons are the source of CMR at high temperatures ($T>T_{\text N1}$).
Polarons are coherent islands of FM spins polarized by coupling to itinerant electrons.
They vary in size from a few to tens of angstroms. 
While polarons could be the source of the resistivity peak and the associated CMR, critical magnetic fluctuations could also produce a similar phenomenon.

\begin{figure}
  \includegraphics[width=0.46\textwidth]{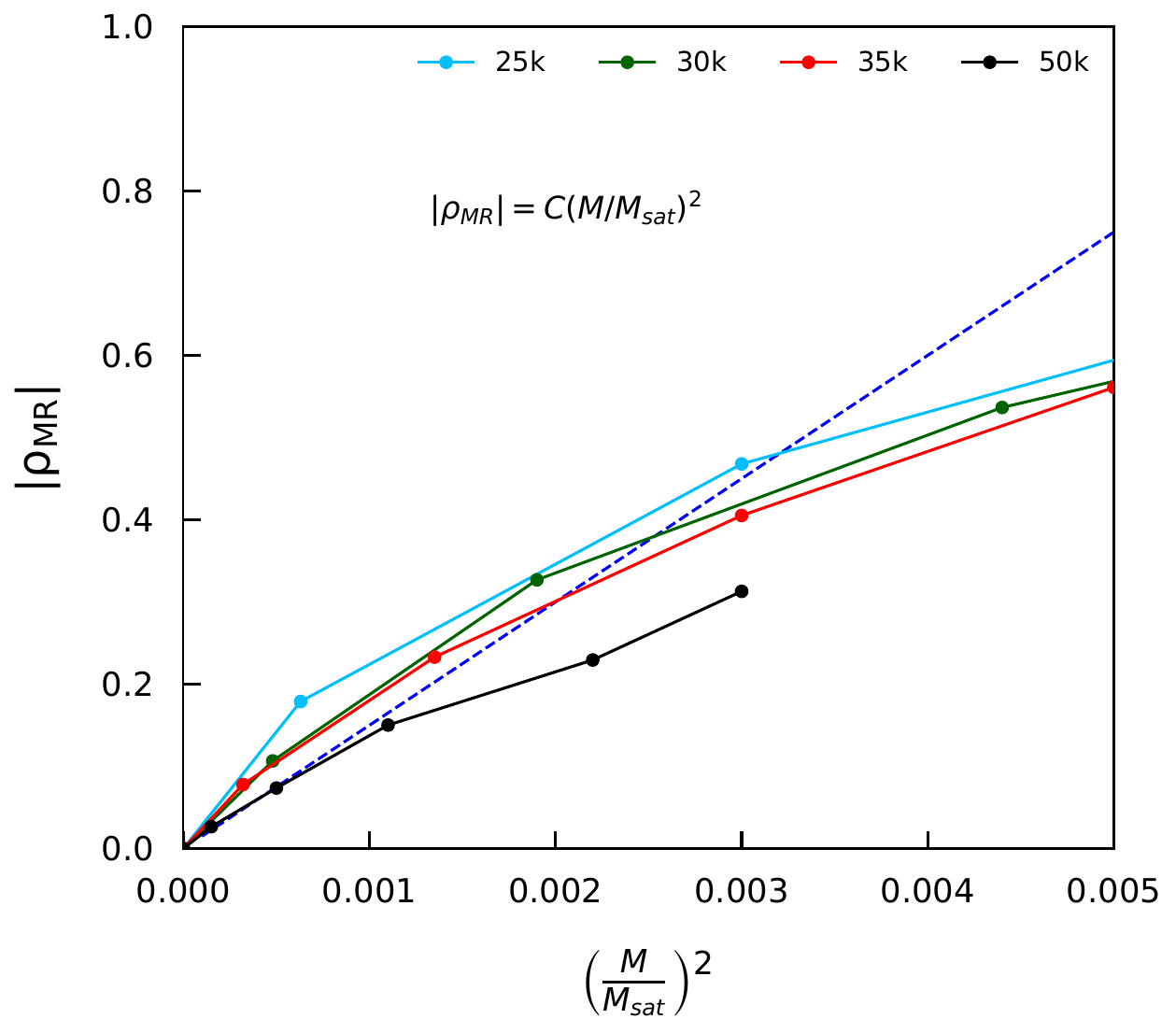}
  \caption{\label{fig:scaling}
 Magnetoresistance vs normalized magnetization squared.}
\end{figure}
\begin{figure}
  \includegraphics[width=0.46\textwidth]{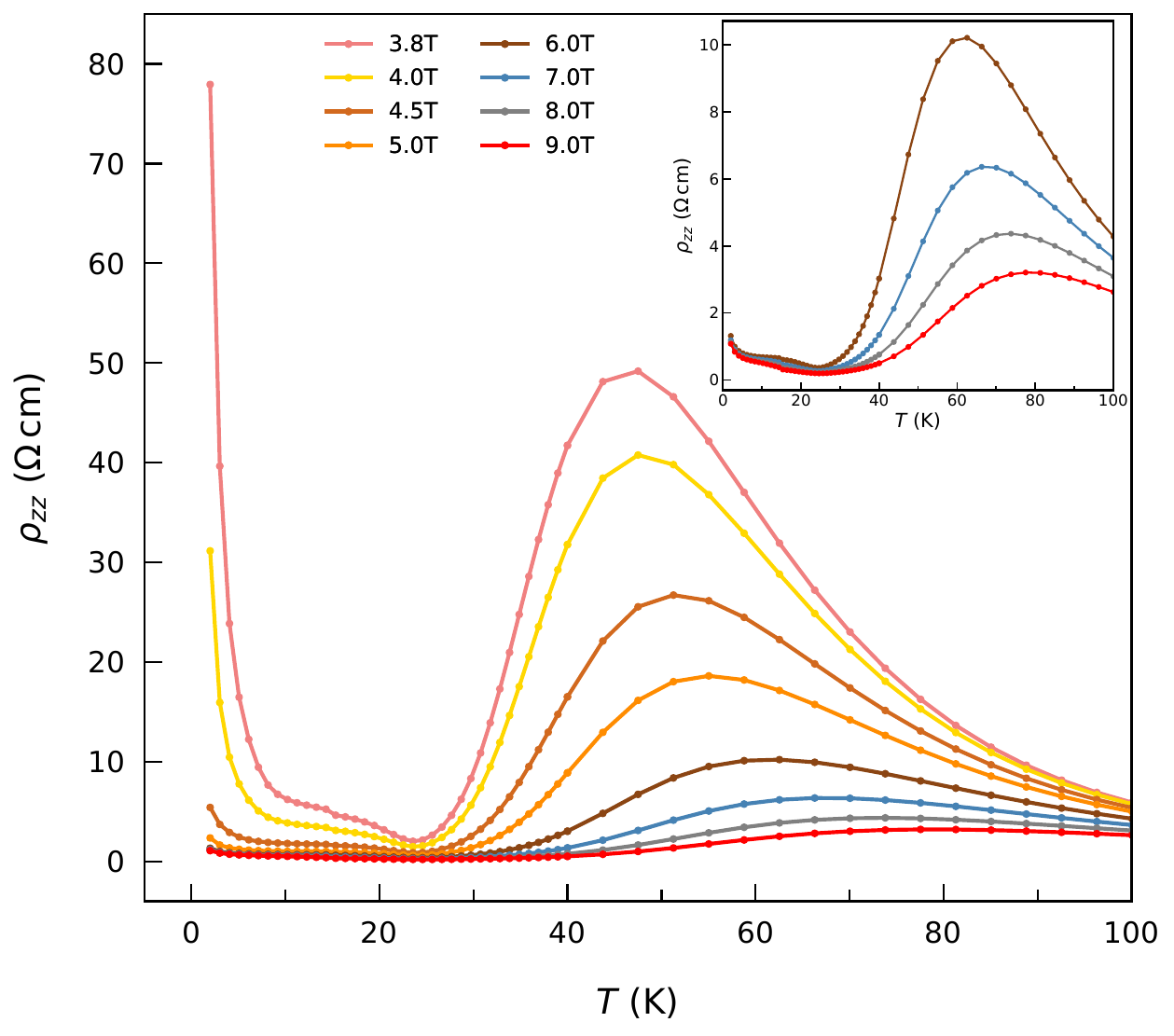}
  \caption{\label{fig:highfield}
 High field $\rho_{zz}$ curves showing the difference in field suppression of resistivity from two types of CMR. The inset highlights the 6–9 T curves, illustrating the difference between the two types of CMR. }
\end{figure}

Theoretically, it is expected that the magnetoresistance scales with magnetization when critical fluctuations are responsible for the CMR.
Such a scaling analysis comes from a Ginzburg-Landau theory, which is explained in Ref.~[40] and references thereof.
In Figure~\ref{fig:scaling}, we show the scaling of low-field magnetoresistance with the square of the normalized magnetization, $\frac{\Delta \rho}{\rho_0}$ (denoted as $\rho_{\text{MR}}$) = $C \left(\frac{M}{M_{\text{sat}}}\right)^2$, where $M_{\text{sat}}$ is the saturation magnetization and $C$ is the scaling factor. 
The scaling results in a value of $C = 150$, which is two orders of magnitude larger than the expected value for critical fluctuations~[40]. 
This suggests the mechanism for magnetoresistance is not due to the suppression of magnetic fluctuations.
A similar deviation in the scaling constant is observed in \ch{EuB6}~[40] and \ch{Eu5In2Sb6}~[11] where percolation of polarons, as their size increases in the magnetic field, is argued to cause the magnetoresistance rather than the suppression of magnetic fluctuations.

\section{\label{sec:RESISTIVITY}Additional Resistivity Data}
In the inset of Fig.~3b in the main text, we show $\rho(2\text{K})$ and $\rho({T^{\rho}_\text{peak}})$ to distinguish the two types of CMR based on how the resistivity peak and the resistivity upturn respond to the magnetic field.
The resistivity data we used to make that plot is presented in Fig.~\ref{fig:highfield}.
These data are obtained from a different sample than the one we used in the main text. 
The salient features of resistivity are the same between these two different samples. 

\begin{figure}
  \includegraphics[width=0.46\textwidth]{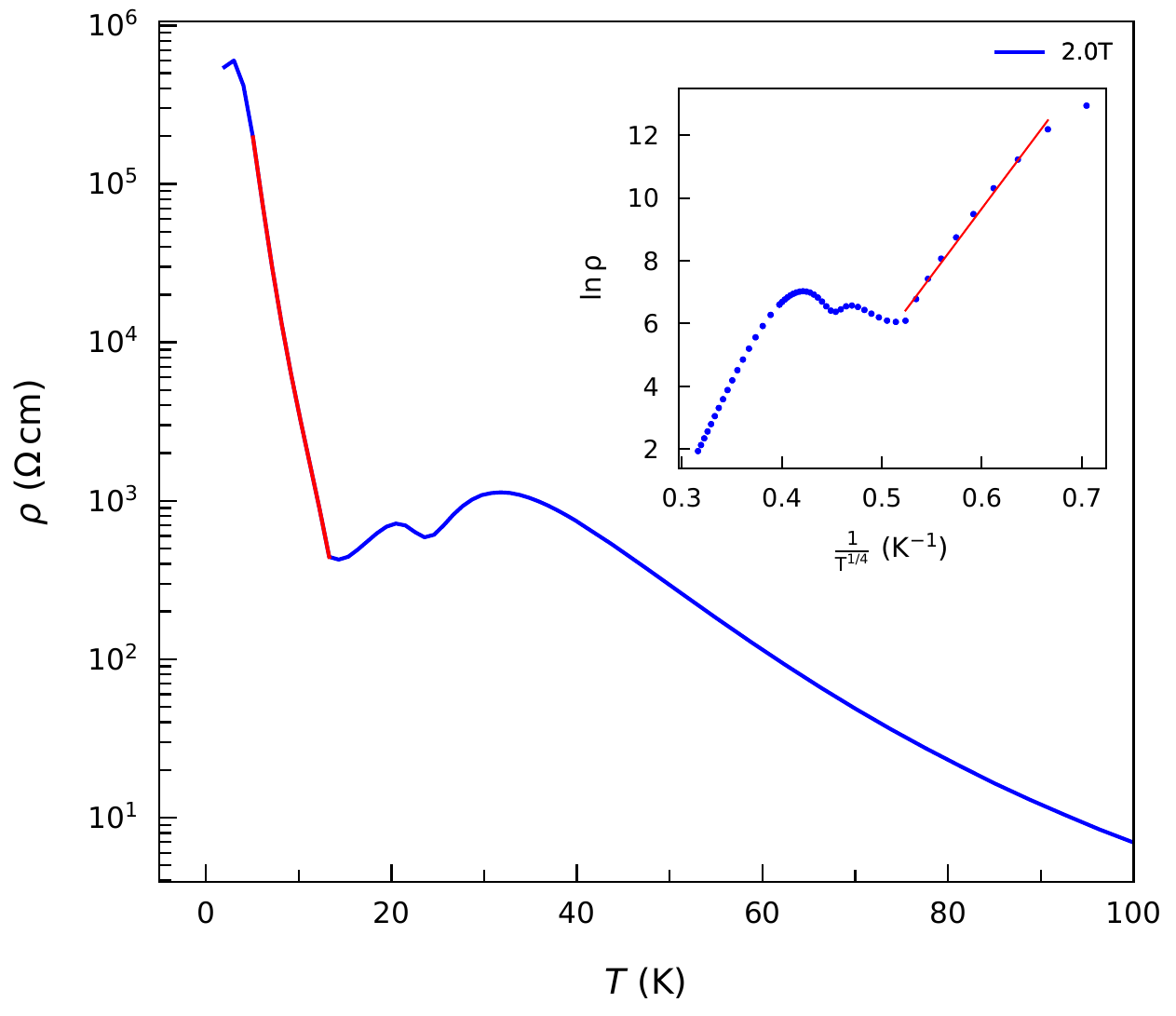}
  \caption{\label{fig:VRH}
 Fits to the variable range hopping model.}
\end{figure}

Figure~\ref{fig:VRH} shows fits to the variable range hopping model, confirming hopping-like transport of impurity states in the sample. 
Note that high-temperature resistivity fits a conventional semiconducting behavior, as shown in Fig.~\ref{fig:S123_GAP}, but the low-temperature data would not fit a single exponential.
Instead, it follows a variable range hopping behavior.
This analysis confirms that the conduction electrons in \ch{Eu5In2As6} come from impurity states with a hopping-type transport through different domains in the sample.
Remarkably, it seems such extrinsic carriers from impurity states are responsible for the sensitive response of the title compound to the magnetic field.

\begin{figure}
  \includegraphics[width=0.46\textwidth]{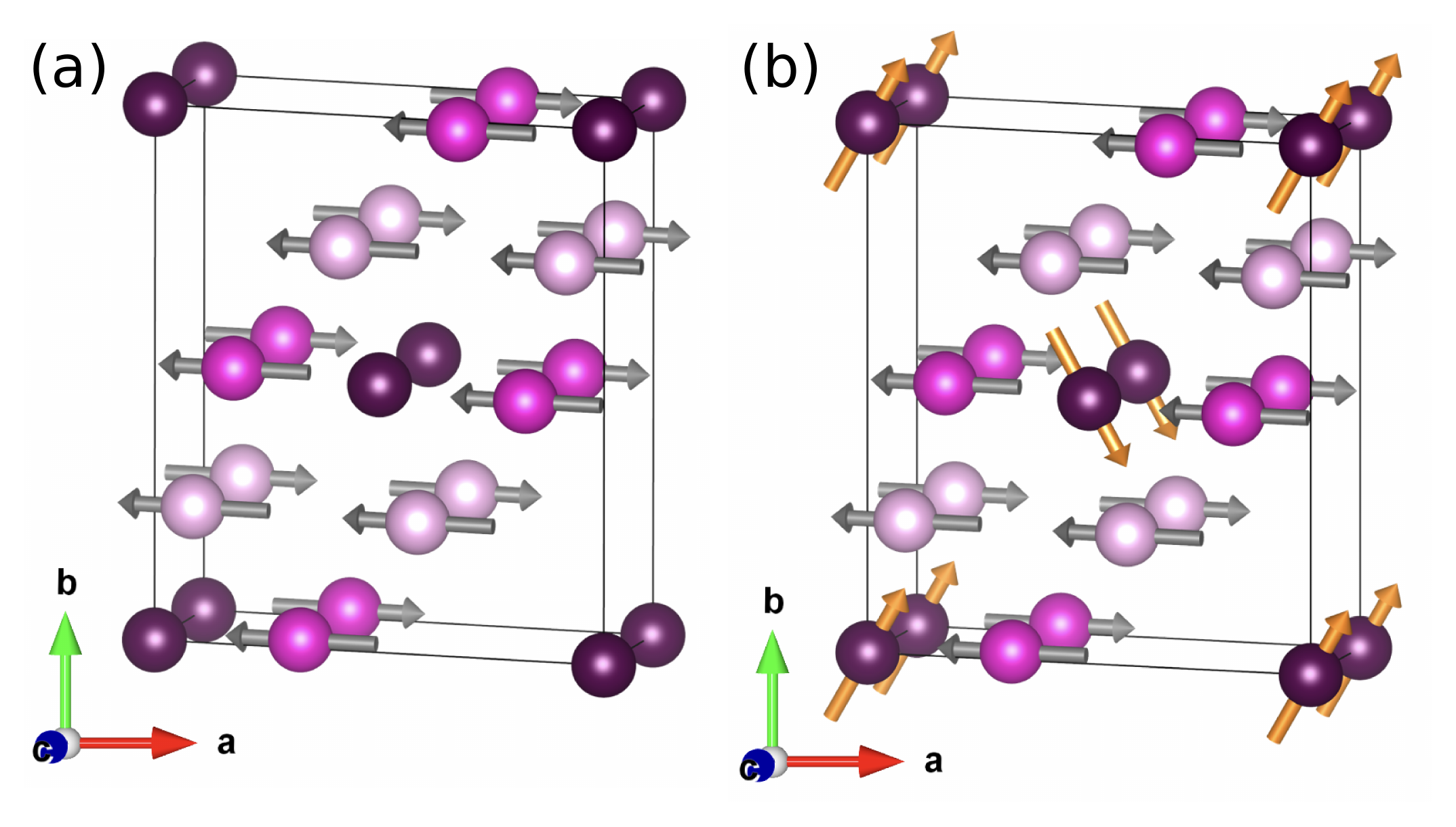}
  \caption{\label{fig:MULTIK}
 The magnetic structure at (a) $T_\text{N2}<T<T_\text{N1}$ and (b) $T<T_{\text{N2}}$ in a scenario where the magnetic order does not phase segregate, instead, the entire sample forms a multi-$\mathbf{k}$ spin structure below $T_{\text{N2}}$. 
}
\end{figure}
\section{\label{sec:NEUTRON}Multi-$k$ scenario}
In the main text, we interpreted the neutron diffraction results in terms of phase separation between the AFM-1 and AFM-0 spin structures shown in Fig.1(e,f).
A different interpretation would be to evoke a multi-$\mathbf{k}$ order below $T_{\text N2}$.
In this case, a multi-$\mathbf{k}$ structure implies both the $\mathbf{k_1}=(0,0,\frac{1}{2})$ and $\mathbf{k_2}=(0,0,0)$ form within the same domain.
As reported in the main text, both $\mathbf{k_1}$ and $\mathbf{k_2}$ order within the $\Gamma_3$ irreducible representation, which both have a finite FM spin component polarized along the $a$-axis.
However, the $\Gamma_3(\psi_1)$ for the $\mathbf{k_1}=(0,0,\frac{1}{2})$ order reverses its polarization along the $c$-axis whereas the $\Gamma_3(\psi_1)$ for the $\mathbf{k_2}=(0,0,0)$ order maintains its polarization along the $c$-axis.
If $\mathbf{k_1}$ and $\mathbf{k_2}$ coexist on the same Eu Wyckoff site, then it would lead to a spatial variation of the magnetization on such a site, which is nonphysical for a localized spin system.
Assuming a multi-$\mathbf{k}$ scenario, one needs to conclude that the $\mathbf{k_1}$ and $\mathbf{k_2}$ orders arise on different Eu sites. 
Considering the small entropy released at $T_{N2}$ and the weak intensity of the $\mathbf{k_2}$ Bragg peaks, we postulate that the Eu3 sites, which have half the occupancy of the Eu1 and Eu2 sites, are the ones ordering with the $\mathbf{k_2}$ wave vector.
Assuming homogeneous total magnetization on all three Eu sites, the spin structure of \ch{Eu5In2As6} is illustrated in Fig.~\ref{fig:MULTIK}a for $T_{\text{N2}}<T<T_{\text{N1}}$ and in Fig.~\ref{fig:MULTIK}b for $T<T_{N2}$.
These spin structures are consistent with the $\chi(T)$ and $M(H)$ curves shown in Figs.~1c,d in the main manuscript.
Specifically, they show FM spins within the $ab$-plans with an easy $a$-axis, and with alternating directions along the $c$-axis.
The Eu3 spins order only below $T_\text{N2}$ in a way that amounts to a FM component along the $a$-axis, explaining the hysteresis observed in the inset of Fig.1(c) at 2~K.

Given the complexity of the multi-$k$ spin structure shown in Fig.~\ref{fig:MULTIK}b and the likelihood of a charge segregation as the mechanism of resistivity upturn, we favor the magnetic phase separation scenario as presented in the main text.



